%% file: main.tex
\documentclass[10pt,twocolumn,letterpaper]{article}

\usepackage{cvpr}

\input{preamble}

\definecolor{cvprblue}{rgb}{0.21,0.49,0.74}
\usepackage[pagebackref,breaklinks,colorlinks,allcolors=cvprblue]{hyperref}

\def\confName{CVPR}
\def\confYear{2026}

\title{Cinematic Audio Source Separation Using Visual Cues}

\author{
Kang Zhang$^1$\footnotemark[1] \quad
Suyeon Lee$^1$\footnotemark[1] \quad
Arda Senocak$^2$\footnotemark[2] \quad
Joon Son Chung$^1$\footnotemark[2]\\
$^1$School of Electrical Engineering, KAIST, 
$^2$Graduate School of Artificial Intelligence, UNIST\\
{\tt\small \{zhangkang, syl4356, joonson\}@kaist.ac.kr, ardasnck@unist.ac.kr} \\
}

\begin{document}
\maketitle
\footnotetext{{\small * Equal contribution,} \, {\small $\dag$ Corresponding authors.}}

\input{sec/0_abstract}

\input{sec/1_intro}

\input{sec/2_related_works}

\input{sec/3_method}

\input{sec/4_AVDnR_dataset}

\input{sec/5_experiments}
\input{sec/6_conclusion}

\section*{Acknowledgements}
This work was supported by Institute of Information \& communications Technology Planning \& Evaluation (IITP) grant funded by the Korean government (RS-2025-02263977, Development of Communication Platform supporting User Anonymization and Finger Spelling-Based Input
Interface for Protecting the Privacy of Deaf Individuals (90\%) and RS-2020-II201336, Artificial Intelligence Graduate School Program (UNIST) (10\%)). 

{
    \small
    \bibliographystyle{ieeenat_fullname}
    \bibliography{main}
}
\input{sec/7_supplementary}

\end{document}

%% file: preamble.tex
\usepackage{bm}
\usepackage{multirow}
\usepackage{booktabs}
\usepackage[table]{xcolor} 
\usepackage{pifont}
\usepackage{algorithm}
\usepackage{algpseudocode}

\usepackage{listings}
\usepackage{xcolor}
\usepackage[symbol]{footmisc}

\definecolor{codegreen}{rgb}{0.0,0.4,0.0}
\definecolor{codegray}{rgb}{0.5,0.5,0.5}
\definecolor{codepurple}{rgb}{0.58,0.0,0.82}
\definecolor{backcolour}{rgb}{0.98,0.98,0.98}

\lstdefinestyle{PyTorchStyle}{
    backgroundcolor=\color{backcolour},   
    commentstyle=\color{codegreen}\textit,
    keywordstyle=\color{blue}\textbf,
    numberstyle=\tiny\color{codegray},
    stringstyle=\color{codepurple},
    basicstyle=\ttfamily\footnotesize,
    breaklines=true,
    showstringspaces=false,
    captionpos=b,
    keywords={for, in, if, else, def, return, class, self, import, from, torch, cat, randn_like, chunk, sigmoid, eval, train, no_grad, view, zeros_like, optimizer},
    language=Python,
}

%% file: sec/0_abstract.tex
\begin{abstract}
Cinematic Audio Source Separation (CASS) aims to decompose mixed film audio into speech, music, and sound effects, enabling applications like dubbing and remastering. Existing CASS approaches are audio-only, overlooking the inherent audio-visual nature of films, where sounds often align with visual cues. We present the first framework for audio-visual CASS (AV-CASS), leveraging visual context to enhance separation quality. Our method formulates CASS as a conditional generative modeling problem using conditional flow matching, enabling multimodal audio source separation. To address the lack of cinematic datasets with isolated sound tracks, we introduce a training data synthesis pipeline that pairs in-the-wild audio and video streams (\eg, facial videos for speech, scene videos for effects) and design a dedicated visual encoder for this dual-stream setup. Trained entirely on synthetic data, our model generalizes effectively to real-world cinematic content and achieves strong performance on synthetic, real-world, and audio-only CASS benchmarks. Code and demo are available at \url{https://cass-flowmatching.github.io}.
\end{abstract}

%% file: sec/1_intro.tex
\section{Introduction}
\label{sec:intro}

\begin{figure}[t]
  \centering
   \includegraphics[width=1\linewidth]{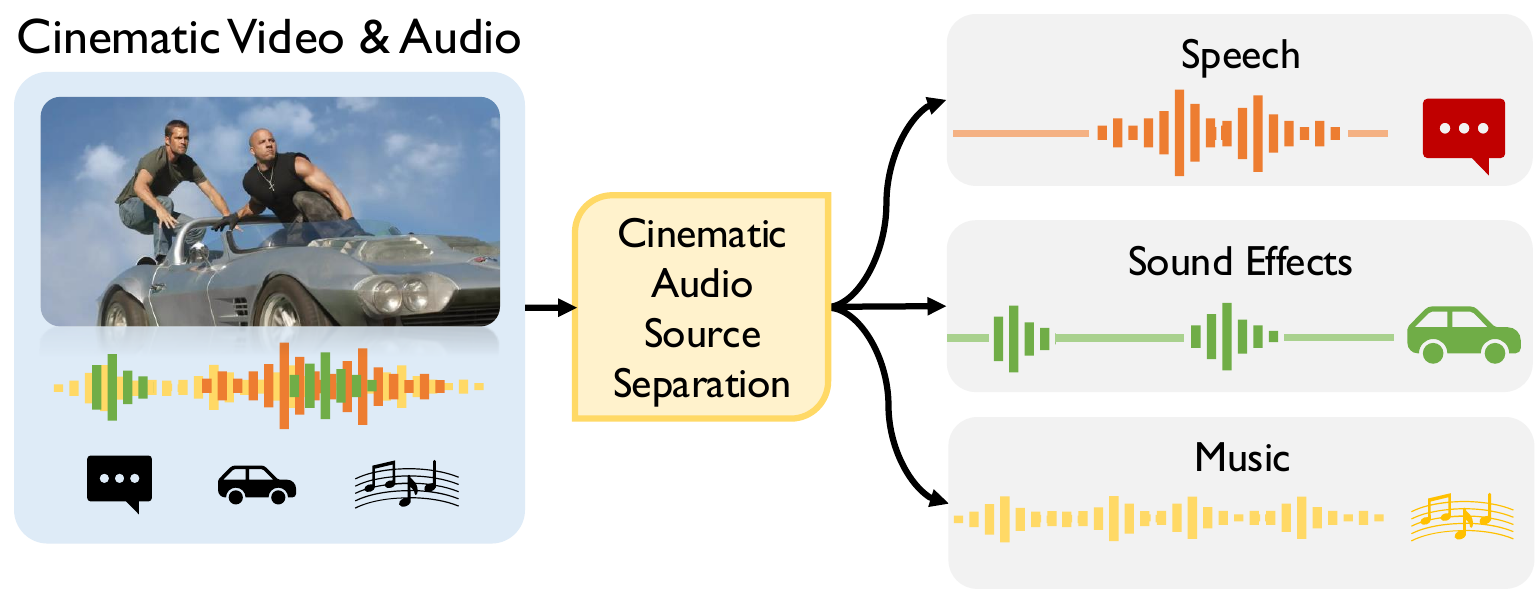}
   \caption{\textbf{Illustration of the Cinematic Audio Source Separation (CASS) task.} The audio stream from a movie is separated into distinct tracks: speech, sound effects, and music.}
   \label{fig:cass_task}
   \vspace{-4mm}
\end{figure}

Cinematic audio is composed of layered sound elements such as speech, music, and sound effects, which collectively enrich storytelling and immersion. The goal of Cinematic Audio Source Separation (CASS) is to separate a mixed movie audio into these three distinct tracks (Fig.~\ref{fig:cass_task}). CASS enables a wide range of applications, including multilingual dubbing, film remastering, audio editing, and accessibility enhancement. As video streaming platforms continue to grow, the need for CASS becomes increasingly important to enable automatic tools that precisely control individual sound components.

While related audio separation tasks, such as speech separation~\cite{Hershey2016deep, Chen2017deep, Dong2017permutation} and music demixing~\cite{defossez2019music, rouard2022hybrid, defossez2021hybrid, mariani2024multisource}, have seen significant progress, CASS remains underexplored. The introduction of the Divide and Remaster (DnR) dataset~\cite{petermann2022cocktail} formalized CASS as a three-way separation problem and initiated new research~\cite{watcharasupat2023bandit, Uhlich-2024, watcharasupat2024remastering}. However, existing methods are purely audio-based and overlook a defining characteristic of film: its audio-visual nature.

In cinematic content, sound is often tightly coupled with visual events. Speech typically co-occurs with lip movements, and sound effects frequently align with object interactions or visual actions. Prior work in audio-visual learning has shown that using visual context, such as facial motion, body movement, and scene composition, significantly improves audio separation and enhancement~\cite{afouras2018conversation, ephrat2018looking, gao2021visualvoice, gao2018learning, tzinis2021into, chen2023iquery}. Yet, to our knowledge, visual information has not been utilized in the context of CASS.

The primary motivation of this paper is to achieve cinematic audio source separation using visual cues. Accordingly, we first formulate CASS as a conditional generative modeling problem. We adopt conditional flow matching~\cite{lipman2023flow} as our generative framework, which has demonstrated strong performance in both image~\cite{esser2024scaling, hu2024latent} and audio~\cite{le2024voicebox, guo2024voiceflow, wang2024frieren, jung2024flowavse} generation tasks. Our model generates clean speech, sound effects, and music tracks from the mixture, conditioned on both audio and visual inputs.

Second, despite the strong potential of visual cues, applying them to CASS remains challenging. A key obstacle is the lack of publicly available audio-visual datasets with clean source tracks, which are difficult -- if not impossible -- to obtain from real films. This raises an important question: \textit{Can we leverage individually available in-the-wild audio-visual data to train an effective audio-visual CASS model?} To address this, we propose a novel training data synthesis strategy. In the absence of paired datasets such as films and their clean sound source tracks, we construct a pipeline that synthetically pairs individually available audio and video sources, \eg, using facial videos for speech and scene video clips for sound effects. This results in a dual-stream video setup that provides controllable, source-specific visual supervision, for which we also design a visual feature extractor to leverage this setup. Importantly, although our model is trained using this synthetic dual-video configuration, we demonstrate that it generalizes effectively to real-world cinematic content without architectural changes, highlighting the practicality and robustness of our training approach.

In summary, our contributions are:
\begin{itemize}
\item We introduce the first framework for audio-visual cinematic audio source separation (AV-CASS), incorporating visual cues to separate speech, sound effects, and music in film audio.
\item We formulate CASS as a generative task using conditional flow matching, enabling flexible and principled modeling of multimodal audio decomposition.
\item We propose a training data synthesis strategy, along with a dedicated visual encoder, to enable training without the need for original source-separated film data.
\item We achieve strong performance on synthetic data, real-world cinematic examples, and standard audio-only CASS benchmarks.
\end{itemize}

%% file: sec/2_related_works.tex
\section{Related Work}
\label{sec:related_work}

\subsection{Cinematic Audio Source Separation}
Cinematic Audio Source Separation (CASS) was formalized by~\cite{petermann2022cocktail, petermann2023tackling}, introducing the Divide and Remaster (DnR) dataset for the separation of speech, sound effects, and music in film audio. BandIt~\cite{watcharasupat2023bandit} applied Band-split RNNs~\cite{luo2023music} to this task, achieving improved performance over earlier methods. More recently, DnRv3~\cite{watcharasupat2024remastering} expanded the dataset with multilingual content and introduced mixing strategies aligned with industrial audio production pipelines. DnR-nonverbal~\cite{hasumi2025dnr} further expands the dataset by incorporating nonverbal sounds, such as laughter and screams.
While these efforts have advanced audio-based CASS, they remain limited to audio-only learning. To our knowledge, no prior work has explored audio-visual approaches for CASS, despite the inherently multimodal nature of cinematic content. This gap largely stems from the difficulty of acquiring datasets with both isolated audio tracks and temporally aligned video. In contrast, we propose the first audio-visual CASS framework and introduce a training data synthesis pipeline that constructs audio-visual training samples from in-the-wild video sources, without requiring ground-truth source audio with film datasets. Our approach enables effective audio-visual learning and generalizes well to real-world cinematic content.

\subsection{Audio-Visual Source Separation}
Incorporating visual information has proven highly effective in sound source separation. Prior work has shown that visual cues such as lip movements align strongly with spoken content~\cite{chung2017lip, ma2022visual}, while facial features provide cross-modal biometric information that enhances speech separation~\cite{afouras2018conversation, ephrat2018looking, owens2018audio, afouras2020self, chung2020facefilter, nagrani2018seeing}. Beyond speech, visual information from instrument motion or class-level appearance cues has also been leveraged to improve music separation~\cite{zhao2018sound, gao2019co, gan2020music, chatterjee2021visual}. More generally, recent works have demonstrated that visual context benefits generic sound separation across a wide range of categories~\cite{gao2018learning, tzinis2021into, tzinis2022audioscopev2, chen2023iquery, takahashi2025mmaudiosep, yu2025dgfnet, cheng2025omnisep, huang2024davis, huang2025davisflow}. Despite these advances, applying audio-visual learning to the complex, multi-source nature of cinematic audio remains  unexplored. In this work, we introduce audio-visual learning to Cinematic Audio Source Separation (CASS). 
Unlike prior AVSS approaches that condition on a single visual cue (e.g., facial motion~\cite{chung2017lip, ma2022visual} or object appearance~\cite{gao2018learning, tzinis2021into, huang2025davisflow}), our framework integrates two complementary visual streams, facial and scene context, derived from the same video, enabling individual-source training while applicable to real-world single-video inputs.

\subsection{Flow Matching Models}
Flow matching~\cite{lipman2023flow, esser2024scaling} has recently gained attention as an efficient alternative to diffusion models, offering faster inference by following shorter and more direct generation trajectories. Recent works have applied flow matching to audio synthesis, separation, and enhancement~\cite{le2024voicebox, mehta2024matcha, guo2024voiceflow, wang2024frieren, yuan2025flowsep, huang2025davisflow, jung2024flowavse}, demonstrating its potential to produce high-quality, natural-sounding outputs. 
Previous non-generative, masking-based separation models often introduce artifacts (\textit{e.g.}, spectral holes) as noted in \cite{yuan2025flowsep}, rendering the output unsuitable for downstream tasks like audio editing. We therefore adopt a generative flow-matching model for the CASS task, which effectively resolves this issue.
To our knowledge, this is the first visually conditioned generative flow-matching approach to CASS, designed to yield perceptually natural and artifact-free separated audio suitable for cinematic production.

%% file: sec/3_method.tex
\begin{figure*}[t]
  \centering
   \includegraphics[width=\linewidth]{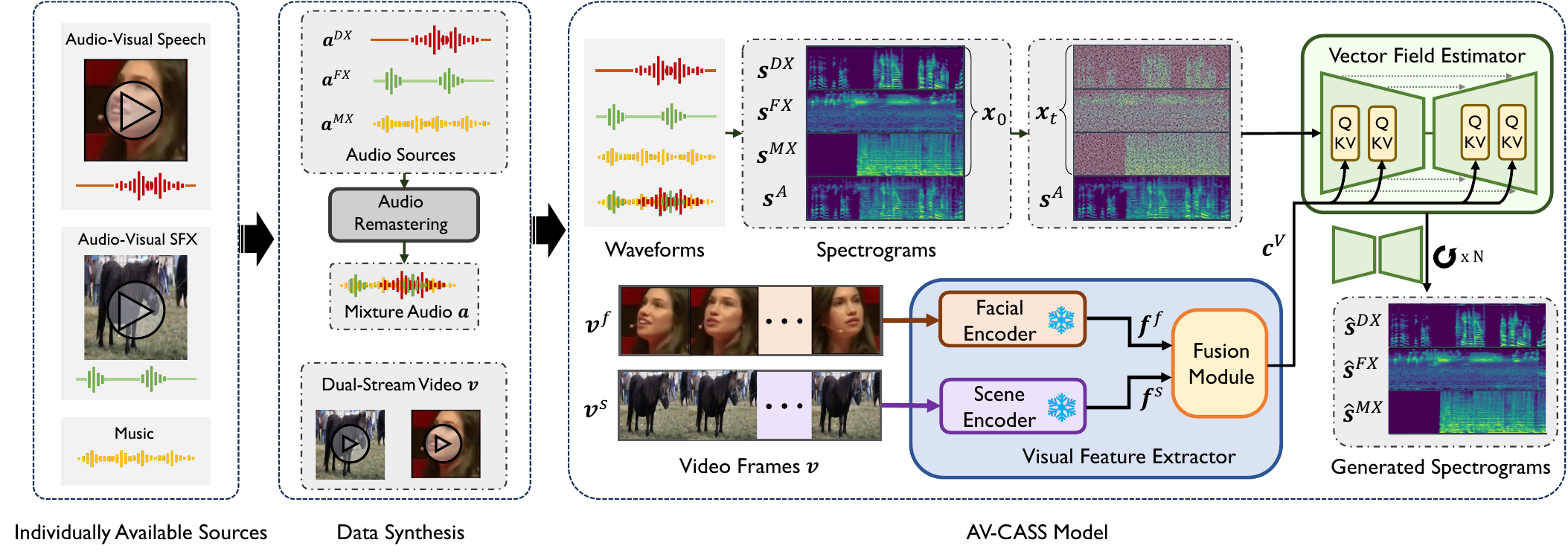}
   \caption{\textbf{Architecture of AV-CASS.} The fusion module integrates visual features from the facial and scene encoders into $\bm{c}^V$, which serves as a conditioning input along with a mixture audio $\bm{s}^A$ for the vector field estimator $\bm{u}_\theta$.
   } 
   \label{fig:method_overview}
   \vspace{-4mm}
\end{figure*}

\section{Methodology}
\label{sec:method}
We propose a framework for audio-visual cinematic audio source separation (AV-CASS). As shown in Fig.~\ref{fig:method_overview}, it consists of a \textbf{Vision Extractor} that generates a fused representation $\bm{c}^V$ from input videos to condition the separation model. Next, a \textbf{flow-based generative model} performs source separation by mapping Gaussian noise to three distinct audio components: speech, sound effects, and music. Finally, to enable training without source-separated film data, we introduce a \textbf{training data synthesis pipeline} that leverages individually available in-the-wild audio and video sources.

\subsection{Problem Setup}
Given a cinematic audio mixture $\bm{a}$ composed of three additive sources, speech $\bm{a}^{DX}$, sound effects $\bm{a}^{FX}$, and music $\bm{a}^{MX}$, the goal of CASS is to recover each source track from the mixture: $\bm{a} = \bm{a}^{DX} + \bm{a}^{FX} + \bm{a}^{MX}$.
We formulate this task as a conditional generation problem, where the model is conditioned not only on the mixture audio but also on visual input $\bm{v}$ from the corresponding video. The video input contains two components: facial frames $\bm{v}^{f}$ and scene frames $\bm{v}^{s}$, which provide complementary cues for speech and sound effects, respectively.

\subsection{Input Data for Training}
To enable training without the need for source-separated cinematic data, we synthesize a training data by combining individual video and audio segments from diverse sources. For each training sample, we: (1) select speech clips from an audio-visual speech dataset to create the speech track $\bm{a}^{DX}$ and the corresponding video as face stream $\bm{v}^{f}$; (2) select sound effect clips from an audio-visual sound dataset to create the sound effects track $\bm{a}^{FX}$ and the corresponding video as scene stream $\bm{v}^{s}$; and (3) select background music clips from a music dataset to create the music track $\bm{a}^{MX}$. These source tracks are mixed to create the input mixture $\bm{a}$, while the video consists of two parallel streams, $\bm{v} = \{\bm{v}^f, \bm{v}^s\}$: a face stream and a scene stream that represent different semantic aspects of the sound. This strategy offers a diverse, controllable, and scalable source of training pairs with ground-truth supervision for all components. Details of the training data synthesis process are provided in Sec.~\ref{sec:data_construct}. All audio waveforms are converted into their corresponding spectrograms where $\bm{s}^A$ denotes the mixture spectrogram, while $\bm{s}^{DX}$, $\bm{s}^{FX}$, and $\bm{s}^{MX}$ represent the individual source spectrograms. The video streams $\bm{v}^f$ and $\bm{v}^s$ are represented as sequences of frames.

\subsection{AV-CASS Model}
Our audio-visual cinematic source separation model (AV-CASS) consists of two main components: (1) visual feature extraction and fusion, and (2) conditional flow matching for source generation.

\subsubsection{Visual Feature Encoding and Fusion}
\label{sec:visual}
We extract visual features from two video streams: facial frames $\bm{v}^f$ and scene frames $\bm{v}^s$, using separate encoders suited to their semantic roles. The facial encoder, adopted from AVDiffuSS~\cite{lee2024seeing}, is designed for lip-synced speech videos. The scene encoder, based on the CAVP model~\cite{luo2023diff}, captures temporally and semantically aligned features of sounding objects and events. Both encoders are frozen during training. The outputs are feature sequences $\bm{f}^f \in \mathbb{R}^{T_f \times D_f}$ and $\bm{f}^s \in \mathbb{R}^{T_s \times D_s}$, where $T_f$ and $T_s$ denote the number of frames, and $D_f$, $D_s$ are feature dimensions. To fuse the representations, we project each stream into a shared feature space of dimension $C$ using separate MLPs:
\begin{equation}
    \bm{f}^f \rightarrow \mathbb{R}^{T_f \times C}, \quad \bm{f}^s \rightarrow \mathbb{R}^{T_s \times C}.
\end{equation}
The resulting features are concatenated along the temporal axis and passed through a final fusion MLP to obtain the visual condition vector:
\begin{equation}
    \bm{c}^V \in \mathbb{R}^{(T_f + T_s) \times C'}.
\end{equation}
This visual condition is used to guide the audio generation process via cross-attention in the U-Net backbone. 

\subsubsection{Flow Matching for Multisource Separation}
We adopt conditional flow matching~\cite{lipman2023flow} to model the conditional joint distribution of source spectrograms $p_1(\bm{x}):=p(\bm{s}^{DX}, \bm{s}^{FX}, \bm{s}^{MX})$ given the mixture spectrograms $\bm{s}^A$ and visual conditioning vector $\bm{c}^V$. Conditional flow matching defines a conditional mapping between the Gaussian noise distribution $\bm x_0\sim \mathcal{N}(\mathbf{0}, \bm{I})$, and the target joint distribution of source spectrograms $\bm{x}_1 \sim p_1(\bm{x})$. 
This mapping defines a time-varying probability density governed by an ordinary differential equation:
\vspace{-1mm}
\begin{equation}
   {\rm d} \bm{x}_t = {\bm u}_\theta(\bm{x}_t, t|\bm{c}) {\rm d} t,
\end{equation}
where ${\bm u}_\theta$ is a vector field estimator representing the gradient of the probability density w.r.t. time $t$ at point $\bm{x}_t$; $\bm{x}_t$ is a point in the probability density space at time $t$; and $\bm{c}$ is conditioning variable includes mixture audio $\bm{s}^A$ and its visual context $\bm{c}^V$.

To construct this mapping, we train ${\bm u}_\theta$ to approximate a reference vector field $\bm{u}_t$, which constructs a probabilistic path between the noise distribution $p_0(\bm{x})$ and the target distribution $p_1(\bm{x})$, conditioned on $\bm{c}$.
The vector field for a noise-data pair $(\bm{x}_0, \bm{x}_1)$ is defined as $\bm{u}_t = \bm{x}_1 - \bm{x}_0$. 
In \cite{lipman2023flow}, given a target distribution sample $\bm x_1$, the data point $\bm{x}_t$ for timestep $t$ on the path is defined as: 
\vspace{-1mm}
\begin{equation} 
\bm{x}_t = (1 - t) \bm{x}_0 + t \bm{x}_1, 
\label{eq:forward} 
\end{equation} 
where $\bm{x}_0 \sim \mathcal{N}(\mathbf{0}, \bm{I})$ and $t \in [0, 1]$. In practice, ${\bm u}_\theta$ is trained to approximate $\bm{u}_t$ by minimizing the following loss:
\begin{equation} 
\mathcal{L} = \mathbb{E}_{t, \pi_1(\bm{x}_1), \pi_0(\bm{x}_0)} \| {\bm u}_\theta(\bm{x}_t, t | \bm{c}) - (\bm{x}_1 - \bm{x}_0)  \|^2_2. 
\label{eq:loss}
\end{equation}
Inspired by \cite{esser2024scaling}, we sample the timestep $t$ from a logit-normal distribution, as this has been shown to enhance generation quality by placing more emphasis on the intermediate timesteps during training.
In practice, we first sample a random variable $z$ from a standard Gaussian distribution $z\sim\mathcal{N}(0,1)$, then map it with a logistic function as follows:
\vspace{-1mm}
\begin{equation}
    t = \frac{1}{1+e^{-z}}, \quad z\sim\mathcal{N}(0,1).
    \label{eq:t_sample}
\end{equation}
For the vector field estimator ${\bm u}_\theta$, we adopt a CNN-based U-Net architecture and three sources are concatenated along the channel dimension to form the input, as commonly done in diffusion-based image generation models~\cite{rombach2022high, ho2020denoising}.

\subsubsection{Inference}
\label{sec:inference} 
\begin{figure}[t]
  \centering
  \includegraphics[width=\linewidth]{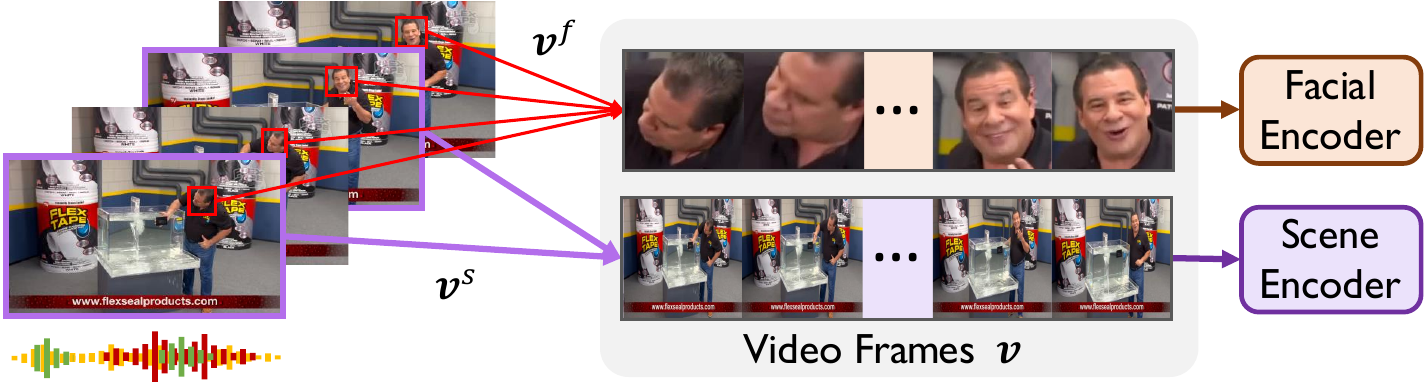}
   \caption{\textbf{Extraction of dual-stream visual inputs from a real-world cinematic video during inference.} Since no architectural changes are required, the AV-CASS model can be used with real-world cinematic videos for inference.
   }
   \label{fig:flextape}
   \vspace{-4mm}
\end{figure}

At inference, the trained conditional flow matching model generates separated source spectrograms from a mixture audio and its associated visual context. As outlined earlier, our training setup uses two video streams for a single audio mixture. Although this differs from the real-world one-video-one-audio setting, our approach remains effective for real-world cinematic content. Since different visual regions (\eg, faces, environments, background elements) correspond to distinct sound sources, as shown in Fig.~\ref{fig:flextape}, we extract the facial regions as one stream for speech cues (processed by the Facial Encoder) and the full scene frames as another for sound effects (processed by the Scene Encoder). This design enables our model to process real-world samples without architectural changes. The model takes as input the mixture spectrogram $\bm{s}^A$ and fused visual condition $\bm{c}^V$ and outputs the separated components: speech, sound effects, and music.

To formulate inference under the conditional flow matching framework, we follow the sampling procedure defined in~\cite{lipman2023flow}. Specifically, we initialize the sample $\bm{x}_0$ as Gaussian noise, $\bm{x}_0 \sim \mathcal{N}(\mathbf{0}, \mathbf{I})$, and iteratively update it along the vector field predicted by vector field estimator $\bm{u}_\theta$, which is conditioned on both the mixture audio and visual context. We apply the forward Euler method to integrate the flow:
\vspace{-1mm}
\begin{equation}
    \bm{x}_{t+\eta} = \bm{x}_t + \eta \, \bm{u}_\theta(\bm{x}_t, t \mid \bm{s}^A, \bm{c}^V),
    \label{eq:euler_sample}
\end{equation}
where $\eta = 1/N$ is the step size and $N$ is the total number of sampling steps. The time variable $t$ progresses from 0 to 1, and at each step, the vector field guides the sample closer to the data distribution of clean source spectrograms. After $N$ steps, the final output $\bm{x}_1$ represents a concatenated output of three spectrograms, each corresponding to a separated source: $\{\hat{s}^{DX}, \hat{s}^{FX}, \hat{s}^{MX}\}$. These spectrograms are converted back to the waveform domain using inverse STFT.

%% file: sec/4_AVDnR_dataset.tex
\section{Training Data Construction Pipeline}
\label{sec:data_construct}
Training an audio-visual CASS model requires synchronized film video and isolated source audio tracks, which are rarely available for real-world cinematic content. To address this, we design a pipeline that synthetically pairs independently sourced audio and video from existing datasets to create realistic training pairs resembling cinematic data. This pipeline produces multimodal training data that mimics cinematic audio-visual patterns while preserving ground-truth alignment across all audio stems and video streams.

\subsection{Individually Available Sources}
We leverage two large-scale audio-visual datasets and one music dataset, each corresponding to one of the three target stems in CASS:

\noindent\textbf{Speech (DX):} We use LRS3~\cite{afouras2018lrs3}, a lip-synchronized video dataset with high-quality speech that may reflect cinematic dialogue through natural prosody and visual expressiveness. 

\noindent\textbf{Sound Effects (FX):} For ambient, object or action-driven sounds, we use VGGSound~\cite{chen2020vggsound}, an audio-visual dataset containing everyday events and objects. Unlike prior works~\cite{petermann2022cocktail, watcharasupat2024remastering} using audio-only FSD50K, VGGSound provides aligned video for learning visually grounded effects.

\noindent\textbf{Music (MX):} Since background music is typically not visually grounded, we follow standard practice and use the FMA dataset~\cite{fma_dataset}, which contains a wide variety of high-quality music.

\subsection{Audio Preprocessing and Stream Synthesis}
To ensure each sample contains a single, uncontaminated source track, we filter VGGSound and FMA with the SMAD model~\cite{hung2022smad}, removing all segments containing speech and music. After filtering, we obtain a total of 152K DX clips, $\sim$ 62K FX clips, and $\sim$ 49K MX samples. 

Following the protocol in DnRv3~\cite{watcharasupat2024remastering}, we synthesize a cinematic audio by using the DX, FX, and MX tracks. For each track, we randomly sample short clips, concatenate them with overlapping transitions, and apply loudness normalization to meet cinematic mastering standards. The resulting tracks are then mixed by addition:
\begin{equation}
    \bm{a}^{A} = \bm{a}^{DX} + \bm{a}^{FX} + \bm{a}^{MX}.
\vspace{-1mm}
\end{equation}
All audio is converted to mono and resampled at 16kHz. Statistics of the resulting training data are in \textit{Appendix}~\ref{app:stats_avdnr_data}.

\subsection{Visual Stream Synthesis}
We extract the video clips corresponding to the DX and FX streams. Using the timestamps of each selected audio segment, we retrieve the facial video aligned with DX (from LRS3) and the scene video aligned with FX (from VGGSound). The MX stream has no associated visual input.

Each training sample contains a mixed audio stream with ground-truth sources and two video streams, \textit{i.e.}, facial and scene, reflecting how speech and sound effects are visually grounded in real films. Although the dataset is synthetically constructed, it enables precise supervision for multimodal learning and aligns well with cinematic audio-visual conventions. As shown in the Sec.~\ref{sec:experiment}, our model trained on this data generalizes effectively to real-world movies, validating our training data pipeline.

%% file: sec/5_experiments.tex
\section{Experiments}
\label{sec:experiment}

\subsection{Experimental Setup}

\noindent\textbf{Baselines.} We compare our method with existing CASS models, including MRX~\cite{petermann2022cocktail} and BandIt~\cite{watcharasupat2023bandit}, as well as musical instrument separation models such as Hybrid Demucs~\cite{defossez2021hybrid}, HT Demucs~\cite{rouard2022hybrid}, and MSDM~\cite{mariani2024multisource}. 
{Beyond audio-only baselines, we also include the audio-visual sound separation model DAVIS-Flow~\cite{huang2025davisflow} which is the current state-of-the-art model for audio-visual sound separation, to assess the contribution of visual conditioning and to highlight the fundamental differences between CASS and generic audio-visual separation tasks.}
All models are trained from scratch on the same dataset, using their original training configurations and appropriately modified to operate under the CASS setting for fair comparison.

\noindent\textbf{Metrics.}
We use Fréchet Audio Distance (FAD)~\cite{kilgour2019fad} and Kullback-Leibler divergence (KL) from AudioLDM~\cite{liu2023audioldm} to measure distributional similarity between generated and real audio. We also report Perceptual Evaluation of Speech Quality (PESQ)~\cite{pesq} for speech and Scale-Invariant Signal-to-Distortion Ratio improvement (SI-SDRi)~\cite{le2019sdr} in dB, following prior works~\cite{petermann2022cocktail, mariani2024multisource}. For FAD, KL, and SI-SDRi, we report averages across all three sources. In addition, we introduce a new metric, \textit{Wrong Placement Ratio (WPR)}, to estimate the proportion of residual or misplaced components from other stems. WPR is computed using PANNs~\cite{kong2020panns}, a pretrained sound event detection model, and reflects stem-level separation quality without ground-truth references; lower values indicate better isolation. Details on metric calculations is in the \textit{Appendix}~\ref{app:metrics}.
\begin{figure*}[!ht]
  \centering
   \includegraphics[width=0.9\linewidth]{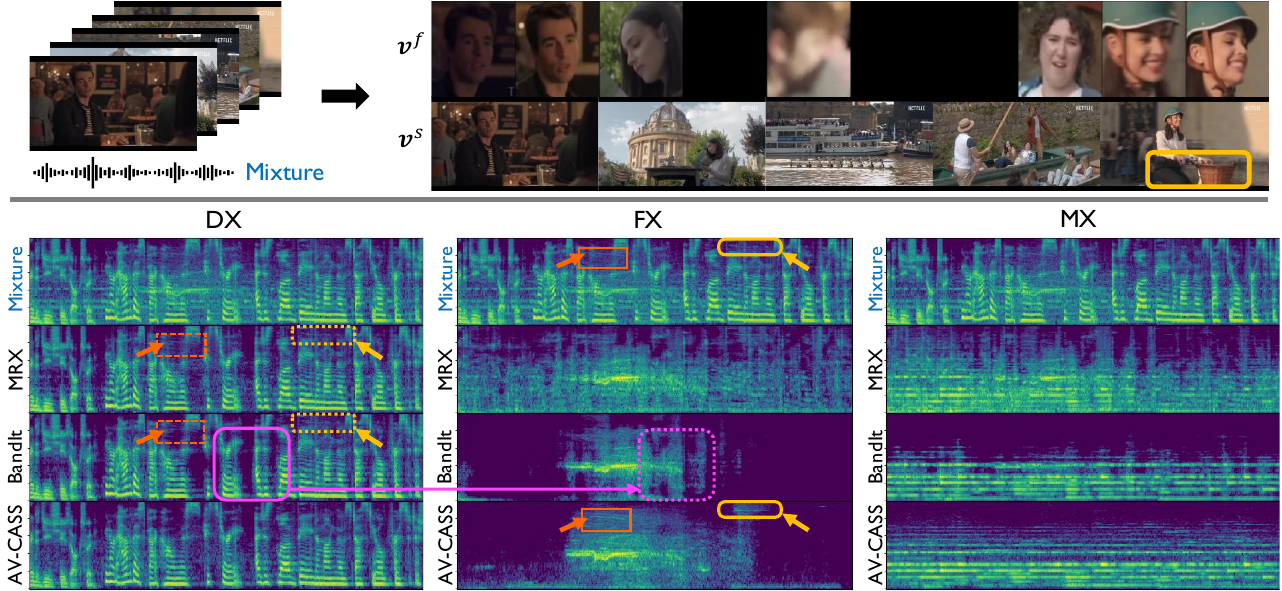}
   \caption{\textbf{Comparison of MRX, BandIt, and AV-CASS on a real-world movie sample.} Input video frames $\bm{v}^f$ and $\bm{v}^s$ are shown at the top, with the input audio spectrogram $\bm{s}^A$ placed for each stem. Yellow boxes highlight the bicycle bell, red boxes indicate cheering, and dotted boxes show elements misplaced in non-target stems. The dotted pink box in BandIt’s FX shows speech artifacts. Best viewed when zoomed in. This sample can also be viewed in the supplementary video.}
   \label{fig:spec_myoxfordyear}
   \vspace{-4mm}
\end{figure*}

\noindent\textbf{Training and Implementation Details.}
Before training the full audio-visual model, we apply audio-denoiser warm-up to stabilize optimization. Specifically, we first train the model using only the audio component of the synthetic audio-visual dataset described in Sec.~\ref{sec:data_construct}. In practice, it improves training stability, accelerates convergence, and prevents the model from prematurely overfitting to visual cues in the early stages of training.

After warm-up, we train the full audio-visual model on the same synthetic dataset. Visual cues from both the facial and scene streams are introduced gradually using zero-initialized convolution layers, following ControlNet~\cite{zhang2023adding} strategy, while the video encoders remain frozen.
This design preserves the stabilized audio representation while allowing the model to gradually integrate visual information.

We use the Adam optimizer~\cite{KingBa15} with $\beta_1=0.9$, $\beta_2=0.999$, a fixed learning rate of $10^{-4}$, and no weight decay. Full audio-visual training runs for 600k steps with a batch size of 8 across four RTX~4090 GPUs. We use 128 sampling steps during evaluation. Additional implementation details and pseudo code are provided in the \textit{Appendix}~\ref{app:exp_detail}.

\subsection{Main Results}

\subsubsection{Evaluation on real-world samples}
\noindent\textbf{Subjective evaluation.}
Generalization to real-world movie samples is critical for CASS models. To evaluate our model, we randomly select clips from the Condensed Movies dataset~\cite{bain2020condensed}, manually verifying that each contains all three target tracks. As this process requires extensive effort, we collected 30 samples. Since ground-truth separation data is unavailable, we conduct a subjective evaluation using mean opinion scores (MOS) through a human study, detailed in the \textit{Appendix}~\ref{app:real-world-mos}. Some of the movie samples used for MOS are available in the supplementary video. We strongly encourage readers to view real-world samples there, along with comparisons to existing CASS methods. The numerical results are provided in Tab.~\ref{tab:mos}. Participants compared and evaluated the separated outputs for each target stem using two criteria: clarity of separation and completeness of target reconstruction. These criteria allow raters to evaluate both how well the model suppresses non-target sources and how fully it preserves the target source without loss of content or quality. Each separation was rated on a 5-point Likert scale, with 1 meaning ``Poor'' and 5 meaning ``Excellent''. As shown in Tab.~\ref{tab:mos}, our model receives higher scores, reflecting user preference and demonstrating the naturalness and sound quality of the outputs. Overall, this result indicates generalization to real-world movie clips.
\input{tables/MOS_movie}
\input{tables/WPR_combined}

\noindent\textbf{Objective evaluation.}
To assess AV-CASS on real-world movies, we report quantitative WPR across DX, FX, and MX stems. As shown in Tab.~\ref{tab:wpr_combined}, our model achieves the lowest WPR for both DX and MX, and delivers competitive performance on FX, indicating strong separation fidelity for dialogue and music. While DAVIS-Flow~\cite{huang2025davisflow} attains a lower WPR on FX, this is expected since it is specifically designed for generic object-centric sound separation; however, it performs poorly on the other tracks. Overall, the results demonstrate that AV-CASS achieves robust cross-track isolation in complex, in-the-wild cinematic audio, supporting the effectiveness of our formulation and synthetic training pipeline.

\noindent\textbf{Qualitative result.}
We also provide a qualitative visual analysis (Fig.~\ref{fig:spec_myoxfordyear}) illustrating residual cross-track components in existing methods, whereas our approach yields cleaner separated tracks. Yellow boxes in $\bm{v}^s$ and the spectrograms highlight the bicycle bell, and red boxes indicate cheering. Dotted boxes show elements incorrectly placed in non-target stems. Other methods retain these effects in the speech track, while ours correctly assigns them to FX. The dotted pink box in BandIt’s FX shows speech artifacts, which AV-CASS avoids. 
This sample and more are shown in supplementary videos.

\input{tables/CASS_trainOurs_testOurs}
\input{tables/CASS_trainOurs_testOurs_WPR}

\subsubsection{Evaluation on AVDnR}
\label{sec:eval_on_avdnr}
While real-world movie audio offers valuable qualitative insights, it does not provide clean ground-truth stems for quantitative evaluation. To enable a more controlled and comprehensive assessment, we construct a fully supervised audio-visual test set, AVDnR, using the same data synthesis pipeline described in Sec.~\ref{sec:data_construct}. We strictly partition all source clips into disjoint training and testing splits to avoid overlap. The final AVDnR benchmark contains 1K audio-visual samples, each 60 seconds long, providing a reliable testbed with complete ground-truth sources.

\noindent\textbf{Objective evaluation.}
We train all methods on our training set and report results on AVDnR in Tab.~\ref{tab:cass_trainOurs_testAVDnR_noMOS}. Methods are grouped into \textit{predictive} and \textit{generative} categories. As commonly observed~\cite{yuan2025flowsep, wen2025promptsep}, predictive models which optimized with reconstruction or SNR-based losses tend to achieve higher SI-SDRi, reflecting their stronger alignment with SNR-based metrics. However, they often produce over-smoothed estimates that limit perceptual fidelity~\cite{saharia2022image, lemercier2023storm}. In contrast, generative models focus on producing realistic samples and therefore excel on perceptual metrics. Within this context, AV-CASS achieves the best FAD, KL, PESQ, and WPR scores across all methods, indicating superior perceptual quality and cleaner cross-track separation.

Notably, AV-CASS outperforms the audio-visual baseline DAVIS-Flow~\cite{huang2025davisflow}, even though both methods use visual information. This suggests that performance in AV-CASS depends not only on using vision, but on the way of providing visual cues. By incorporating facial and scene streams in a dual-stream setup and adopting a multi-source formulation, AV-CASS receives source-specific visual context that is not available in the single-target design of DAVIS-Flow, leading to more consistent and reliable track disentanglement.
Detailed per-track metrics are in the \textit{Appendix}~\ref{app:detailed_metrics}.

\noindent\textbf{Subjective evaluation.}
Since objective scores do not always capture perceived audio quality, especially for generative models that may introduce realistic details beyond the reference, we additionally conduct a user study on AVDnR. As shown in Tab.~\ref{tab:cass_MOS_only}, AV-CASS achieves the highest MOS among all methods. This perceptual preference aligns with our strong FAD, KL, and PESQ performance (Tab.~\ref{tab:cass_trainOurs_testAVDnR_noMOS}), indicating that AV-CASS produces outputs that listeners consistently find clearer and more natural. Together, these results confirm that the perceptual advantages of our model extend beyond objective measures and translate directly into improved user experience.

\begin{figure}[t]
\vspace{-4mm}
  \centering
   \includegraphics[width=0.95\linewidth]{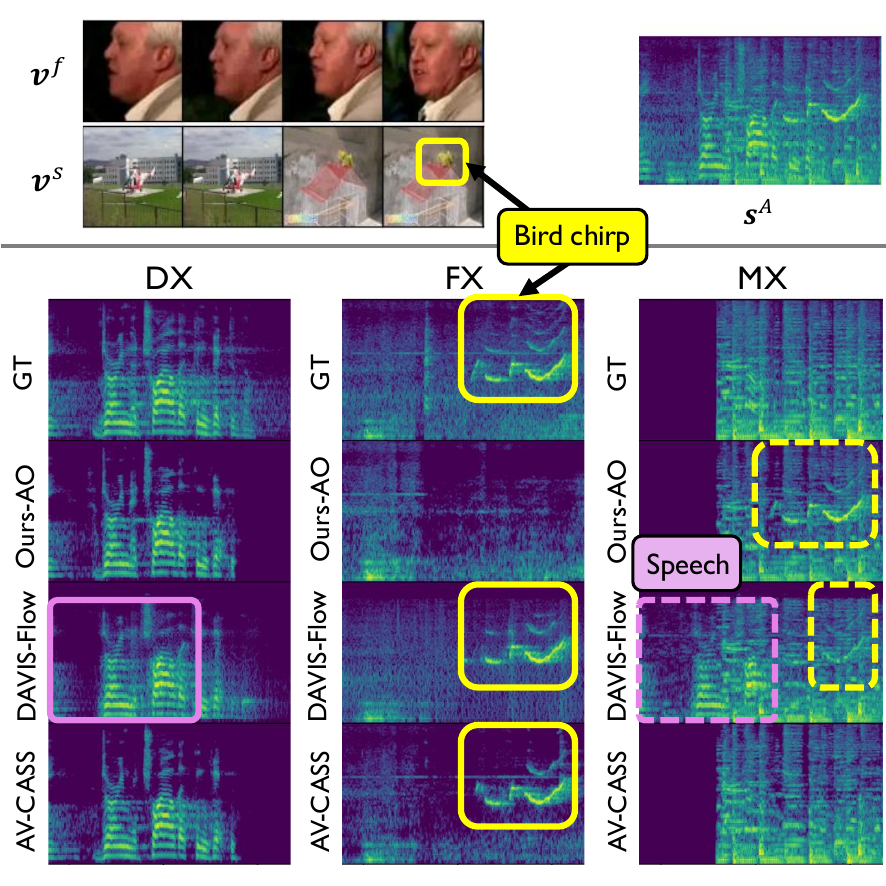}
   \vspace{-2mm}
   \caption{\textbf{Comparison of our audio-only model (Ours-AO), DAVIS-Flow~\cite{huang2025davisflow}, and our audio-visual model (AV-CASS) on a clip from the AVDnR testset.} The input video frames and the GT audio spectrograms are shown at the top. Yellow boxes highlight the bird chirping present in $\bm{v}^s$ and the FX tracks. Dotted boxes indicate misplaced segments. Better viewed when zoomed in.}
   \vspace{-4mm}
   \label{fig:spec_davis}
\end{figure}

\noindent\textbf{Qualitative results.}
Fig.~\ref{fig:spec_davis} compares separated spectrograms from our audio-only model (Ours-AO), DAVIS-Flow~\cite{huang2025davisflow}, and our audio-visual model (AV-CASS). As the yellow boxes indicate, both audio-visual models correctly separate bird chirping sound into FX, guided by the bird visible in $\bm{v}^s$. In contrast, Ours-AO incorrectly places the chirping sound in the MX track. This shows that visual cues can guide correct source separation, resolving ambiguity about where the sound came from.
While DAVIS-Flow separates FX correctly, it fails to isolate MX clearly because its design requires a corresponding visual input for every target source.
This comparison highlights that AV-CASS effectively exploits visual information. Taken together with the results on real-world samples, the gap between AV-CASS and DAVIS-Flow further confirms that the CASS task differs from generic audio-visual sound separation and requires specialized architectures tailored to its multi-source separation demands, such as ours. 
Additional examples are shown in \textit{Appendix}~\ref{app:supple_additional_results}.

\input{tables/CASS_trainOurs_testALL}
\subsubsection{Extendability towards audio-only separation}
As discussed earlier, the CASS task is traditionally defined in the audio-only domain, with standard benchmarks such as DnRv2~\cite{petermann2023tackling} and DnRv3~\cite{watcharasupat2024remastering}. Although our model is designed for audio-visual separation, it can also operate in an audio-only setting, \eg, when video frames are unavailable, by removing the visual encoder and cross-attention blocks in the U-Net. To evaluate this configuration, we train the audio-only variant on our dataset (as in previous experiments) and compare it with other methods on standard audio-only benchmarks and AVDnR. As shown in Tab.~\ref{tab:cass_trainOurs_testAVDnR_noMOS} and Tab.~\ref{tab:cass_trainOurs_testALL}, our method is competitive with models specialized for audio-only CASS while showing clear superiority in FAD, indicating more natural and realistic outputs. Most importantly, our method achieves better WPR performance, indicating cleaner separation with less contamination. These results highlight the flexibility of our approach to support single-modality setups, though its primary scope remains solving CASS from an audio-visual perspective.

\subsection{Ablation Study}
\label{subsec:ablation_study}

\noindent\textbf{Analysis on visual streams.} We study the impact of visual stream by ablating facial and scene video streams. As shown in Tab.~\ref{tab:singlevid_vs_bothvid}, adding either stream improves performance over the audio-only baseline, while using both yields the best results across all metrics, highlighting their complementary benefits and justifying our design choice for AV-CASS. We further analyze how each type of visual input affects the misplacement rates in Tab.~\ref{tab:visual_ablation_wpr_per_stem}. DX and FX columns clearly show that the corresponding visual inputs minimize misplaced segments: the lowest DX WPR is achieved with the facial stream, and the lowest FX WPR with the scene stream. By utilizing both visual cues, our model achieves the most balanced WPR performance across all stems and thus improves the overall perceptual performance. This fine-grained misplacement analysis provides additional evidence that visual cues not only improve signal quality metrics, but also substantially reduce semantic cross-contamination between tracks, a crucial property for practical cinematic audio applications.
Taken together with the results in Tab.~\ref{tab:singlevid_vs_bothvid}, these findings confirm that combining both visual streams offers the most consistent and reliable separation, yielding strong performance across metrics while minimizing residual cross-track contamination.

\input{tables/vid1_vs_vid2}

\input{tables/Visual_input_ablation_WPRs}

\begin{figure}[t]
  \centering
   \includegraphics[width=\linewidth]{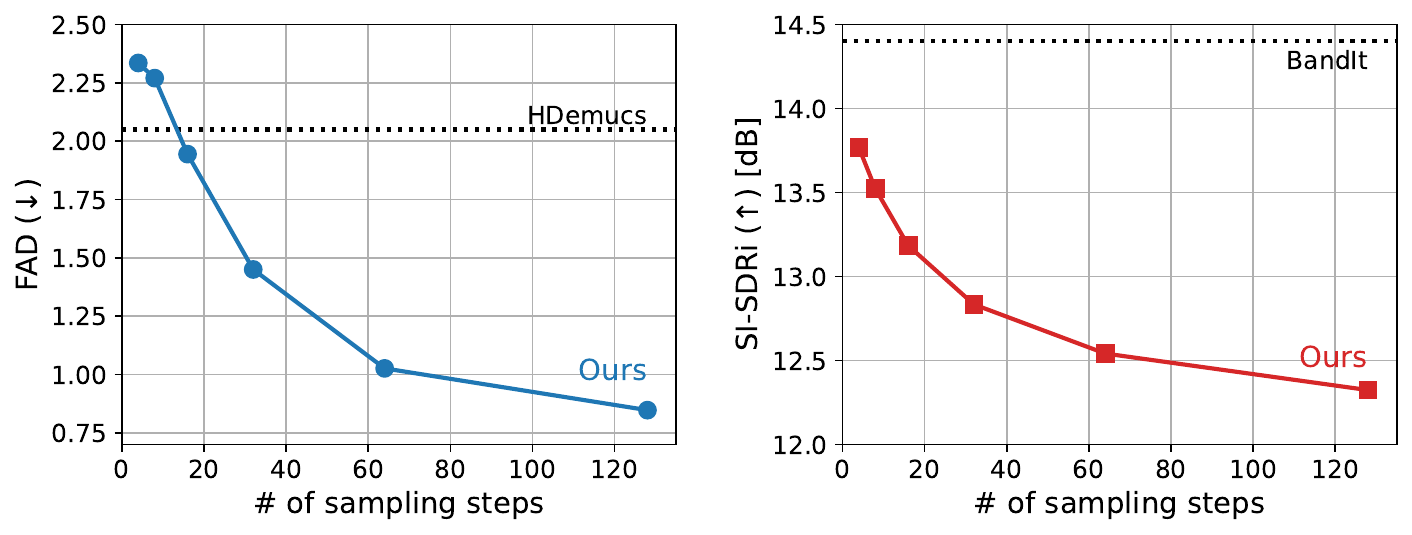}
   \vspace{-6mm}
   \caption{\textbf{Impact of the number of sampling steps.} Dotted lines denote Hybrid Demucs (second-best FAD) on the left and BandIt (highest SI-SDRi) on the right.}
   \label{fig:abl_sampling_steps}
   \vspace{-4mm}
\end{figure}
\noindent\textbf{Impact of sampling step.} 
Since we formulate CASS as a generative task, an important factor is the number of sampling steps $N$, which directly impacts performance. Fig.~\ref{fig:abl_sampling_steps} shows FAD and SI-SDRi results on AVDnR as we vary $N \in \{4, 8, 16, 32, 64, 128\}$. FAD consistently improves with more steps, indicating better perceptual quality. Also, our model with only 32 steps already surpasses the second-best method, Hybrid Demucs~\cite{defossez2021hybrid}, while also improving SI-SDRi. For our experiments, we use 128 steps to maximize perceptual quality, though $N$ can be adjusted depending on the target metric.

%% file: tables/MOS_movie.tex
\begin{table}[t]
  \centering
\resizebox{0.8\linewidth}{!}{ 
\setlength{\tabcolsep}{3pt}
  \begin{tabular}{lccc}
    \toprule
    \textbf{Method} &  MRX~\cite{petermann2022cocktail} & BandIt~\cite{watcharasupat2023bandit} & \textbf{AV-CASS (Ours)} \\
    \midrule
    \textbf{MOS} ($\uparrow$) & 2.55 $\pm$ 0.10 & 3.78 $\pm$ 0.10 & \textbf{4.13 $\pm$ 0.09}  \\ 
    \bottomrule
  \end{tabular}
  }
  \vspace{-2mm}
  \caption{\textbf{MOS results of CASS models on real-world samples.} The scores are computed based on 95\% confidence intervals.} 
  \vspace{-6mm}
  \label{tab:mos}
\end{table}

%% file: tables/WPR_combined.tex
\begin{table}[t]
\centering
\resizebox{0.73\linewidth}{!}{
\setlength{\tabcolsep}{10pt}
\begin{tabular}{lccc}
\toprule
\textbf{Method}   & \textbf{DX} ($\downarrow$)  & \textbf{FX} ($\downarrow$) & \textbf{MX} ($\downarrow$) \\
\midrule
MRX~\cite{petermann2022cocktail}  & 2.63 & 33.72 & 6.01 \\
BandIt~\cite{watcharasupat2023bandit}  & \underline{0.60} & 22.41 & \underline{0.35} \\
DAVIS-Flow~\cite{huang2025davisflow} & 5.88 & \textbf{14.58} & 35.94 \\
\textbf{AV-CASS (Ours)}  & \textbf{0.46} & \underline{19.81} & \textbf{0.32} \\ 
\bottomrule
\end{tabular}
}
\vspace{-2mm}
\caption{\textbf{Wrong Placement Ratio (WPR [\%]) on real-world samples.} Lower is better.}
\label{tab:wpr_combined}
\vspace{-4mm}
\end{table}

%% file: tables/CASS_trainOurs_testOurs.tex
\begin{table}[t]
\centering
\resizebox{1.0\linewidth}{!}{
\setlength{\tabcolsep}{2pt}
\begin{tabular}{lcccccc}
\toprule
  {\textbf{Method}} & \textbf{A-V} & \textbf{FAD ($\downarrow$)} & \textbf{KL ($\downarrow$)} & \textbf{SI-SDRi ($\uparrow$)}  & \textbf{PESQ ($\uparrow$)} & \textbf{WPR ($\downarrow$)}  \\
\midrule
\multicolumn{7}{l}{\cellcolor{lightgray!25}\textit{Predictive Model}} \\
Hybrid Demucs~\cite{defossez2021hybrid} & \ding{55} & 2.05 & \underline{1.03} & \underline{13.57} & \underline{2.16} & 5.24 \\
HT Demucs~\cite{rouard2022hybrid}       & \ding{55} & 2.08 & 1.06            & 13.41            & 2.06 & 9.23 \\
MRX~\cite{petermann2022cocktail}        & \ding{55} & 3.47 & 1.67            & 10.60            & 1.89 & 14.91\\
BandIt~\cite{watcharasupat2023bandit}   & \ding{55} & 2.15 & 1.14            & \textbf{14.40}   & 2.15 & 4.65\\
\midrule
\multicolumn{7}{l}{\cellcolor{lightgray!25}\textit{Generative Model}} \\
MSDM~\cite{mariani2024multisource}      & \ding{55} & 2.90 & 2.90            & 11.63            & 2.12 & 5.65 \\
DAVIS-Flow~\cite{huang2025davisflow}    & \ding{51} & 5.94 & 1.64 & 9.25 & 1.96 & 12.14 \\
\textbf{AV-CASS (Ours)}                 & \ding{51} & \textbf{0.84} & \textbf{0.93} & 12.32 & \textbf{2.26} & \textbf{1.84} \\
\midrule
\textcolor{gray}{Ours (Audio-only)}     & \ding{55} & \underline{\textcolor{gray}{1.63}} & \textcolor{gray}{1.15} & \textcolor{gray}{12.23} & \textcolor{gray}{2.08} & \underline{\textcolor{gray}{2.01}} \\
\bottomrule
\end{tabular}
}
\vspace{-2mm}
\caption{\textbf{Results on AVDnR dataset (objective scores).} All models are trained on our training data. A-V indicates audio-visual.}
\vspace{-2mm}
\label{tab:cass_trainOurs_testAVDnR_noMOS}
\end{table}

%% file: tables/CASS_trainOurs_testOurs_WPR.tex
\begin{table}[t]
\centering
\resizebox{1.0\linewidth}{!}{
\setlength{\tabcolsep}{3pt}
\begin{tabular}{lccccccc}
\toprule
\textbf{Method} 
& H-Demucs 
& HT Demucs 
& MRX 
& BandIt 
& MSDM 
& \textbf{Ours} \\
\midrule
\textbf{MOS ($\uparrow$)} 
& \underline{3.14 \footnotesize{$\pm$ 0.15}}
& 3.01 \footnotesize{$\pm$ 0.14}
& 1.90 \footnotesize{$\pm$ 0.13}
& 3.12 \footnotesize{$\pm$ 0.14}
& 2.79 \footnotesize{$\pm$ 0.14}
& \textbf{3.90 \footnotesize{$\pm$ 0.13}}\\
\bottomrule
\end{tabular}
}
\vspace{-2mm}
\caption{\textbf{MOS results on AVDnR dataset.}}
\label{tab:cass_MOS_only}
\vspace{-4mm}
\end{table}

%% file: tables/CASS_trainOurs_testALL.tex
\begin{table}[t]
\centering
\resizebox{\linewidth}{!}{
\setlength{\tabcolsep}{3pt}
\begin{tabular}{clccccc}
\toprule
&  & \textbf{FAD ($\downarrow$)} & \textbf{KL ($\downarrow$)} & \textbf{SI-SDRi ($\uparrow$)} & \textbf{PESQ ($\uparrow$)} & \textbf{WPR ($\downarrow$)} \\
\midrule
\multirow{6}{*}{\rotatebox[origin=c]{90}{\textbf{DnRv2~\cite{petermann2022cocktail}}}}
    & Hybrid Demucs~\cite{defossez2021hybrid} & 3.66 & 1.47 & \textbf{9.19} & \underline{2.03} & 4.77 \\
    & HT Demucs~\cite{rouard2022hybrid} & 3.72 & \underline{1.37} & 8.59 & 1.95 & 9.14 \\
    & MRX~\cite{petermann2022cocktail} & 5.01 & 1.77 & 7.48 & 1.78 & 16.56 \\
    & BandIt~\cite{watcharasupat2023bandit} & \underline{2.75} & 1.55 & 7.68 & 1.97 & \underline{3.67} \\
    & MSDM~\cite{mariani2024multisource} & 6.25 & 1.54 & \underline{9.09} & \textbf{2.07} & 5.33 \\
    \rowcolor{lightgray!25}
    \cellcolor{white} & \textbf{Ours (Audio-only)} & \textbf{1.95} & \textbf{1.33} & 8.10 & 1.93 & \textbf{2.13} \\ 
\midrule
\multirow{6}{*}{\rotatebox[origin=c]{90}{\textbf{DnRv3~\cite{watcharasupat2024remastering}}}}
    & Hybrid Demucs~\cite{defossez2021hybrid} & \underline{3.12} & 1.68 & \textbf{10.62} & \textbf{1.89} & 3.78 \\
    & HT Demucs~\cite{rouard2022hybrid} & 3.17 & \textbf{1.59} & \underline{9.92} & 1.83 & 8.49 \\
    & MRX~\cite{petermann2022cocktail} & 5.02 & 2.26 & 8.94 & 1.65 & 16.64 \\
    & BandIt~\cite{watcharasupat2023bandit} & 4.79 & 1.98 & 9.14 & \underline{1.86} & \underline{3.53} \\
    & MSDM~\cite{mariani2024multisource} & 5.51 & 1.75 & 9.19 & 1.65 & 4.49 \\
    \rowcolor{lightgray!25}
    \cellcolor{white} & \textbf{Ours (Audio-only)} & \textbf{2.62} & \underline{1.66} & 9.36 & \underline{1.86} & \textbf{1.91} \\ 
\bottomrule
\end{tabular}
}
\caption{\textbf{Audio-only CASS results.} Metrics are averaged across three sources, except PESQ, which is evaluated only on the speech source. All models are trained on our dataset.}
\vspace{-4mm}
\label{tab:cass_trainOurs_testALL}
\end{table}

%% file: tables/vid1_vs_vid2.tex
\begin{table}[t]
  \centering
  \resizebox{\linewidth}{!}{
  \begin{tabular}{lcc|cccc}
    \toprule
    \textbf{Method} & $\bm{v}^f$ & $\bm{v}^s$ & \textbf{FAD} ($\downarrow$) & \textbf{KL} ($\downarrow$) & \textbf{SI-SDRi ($\uparrow$)} & \textbf{PESQ} ($\uparrow$)\\
    \midrule
    Audio-only & \ding{55} & \ding{55} & {1.63} & {1.15} & {12.23} & {2.08} \\
     + Facial stream  & \ding{51} & \ding{55}     & 0.91 & 1.00 & 12.13  & 2.21 \\
     + Scene stream  & \ding{55} & \ding{51}      & 0.87 & 1.00 & 12.27  & 2.24 \\
    \midrule
     \textbf{+ Both (Ours)}  & \ding{51} & \ding{51}      & \textbf{0.84}  & \textbf{0.93} & \textbf{12.32}  & \textbf{2.26} \\
    \bottomrule
    \end{tabular}
  }
  \vspace{-2mm}
  \caption{\textbf{Ablation results on visual streams.}}
  \label{tab:singlevid_vs_bothvid}
  \vspace{-2mm}
\end{table}

%% file: tables/Visual_input_ablation_WPRs.tex
\begin{table}[t]
    \centering
    \resizebox{0.8\linewidth}{!}{
    \begin{tabular}{lcc|ccc}
        \toprule
        \textbf{Method} & $\bm{v}^f$ & $\bm{v}^s$ & \textbf{DX} & \textbf{FX} & \textbf{MX} \\
        \midrule
        Audio-only & \ding{55} & \ding{55} & 0.0265 & 4.6507 & 1.3443 \\
        + Facial stream & \ding{51} & \ding{55} & \textbf{0.0164} & 4.3310 & \textbf{1.2028} \\
        + Scene stream & \ding{55} & \ding{51} & 0.0289 & \textbf{4.0939} & 1.2282 \\
        \midrule
        \textbf{+ Both (Ours)} & \ding{51} & \ding{51} & \underline{0.0255} & \underline{4.2854} & \underline{1.2077} \\
        \bottomrule
    \end{tabular}
    }
    \vspace{-2mm}
    \caption{\textbf{
    Ablation results on visual streams with Wrong Placement Ratio (WPR~[\%]) for each stem.} Lower is better.}
    \label{tab:visual_ablation_wpr_per_stem}
    \vspace{-2mm}
\end{table}

%% file: sec/6_conclusion.tex
\section{Conclusion}
In this work, we present the first audio-visual framework for cinematic audio source separation (CASS). By shifting from predictive to conditional generative modeling and leveraging multimodal cues, our method delivers high-quality, perceptually realistic separation. We develop a synthetic training pipeline that pairs in-the-wild audio and video, enabling training without cinematic datasets containing clean separated tracks. Trained solely on synthetic data, AV-CASS generalizes seamlessly to real cinematic content. Extensive experiments demonstrate its effectiveness on synthetic benchmark, real-world movie samples, and even standard audio-only CASS benchmarks. These results highlight the potential of an audio-visual perspective for building scalable, generalizable CASS models.

%% file: sec/7_supplementary.tex
\clearpage

\renewcommand{\thepage}{S\arabic{page}}
\setcounter{page}{1}

\newpage

\twocolumn[{
\begin{center}
\section*{\LARGE Supplementary Material}
\vspace{1em}
\end{center}
}]

\appendix

\section{Experimental Details}
\label{app:exp_detail}

\subsection{Implementation Details}
Input log-magnitude spectrograms are computed from 16 kHz audio using a 1024-point FFT, a 64 ms window, and a 256-sample hop size, corresponding to 8.192-second clips.
For the visual modality, video streams are sampled at 4 FPS for the scene stream ($\bm{v}^s$) and 25 FPS for the facial stream ($\bm{v}^f$). All frames are resized to $224\times224$ pixels before being fed into their respective encoders.

The facial encoder is adopted from AVDiffuSS~\cite{lee2024seeing}, pretrained on VoxCeleb2~\cite{chung2018voxceleb2} for the audio-visual speech separation task.
The scene encoder follows the design of CAVP~\cite{luo2023diff}, pretrained on the VGGSound dataset via contrastive audio-video alignment.
Both visual encoders are frozen during training, while the fusion module is jointly optimized with the vector field estimator $\bm{u}_\theta$ to obtain a unified multimodal representation $\bm{c}^V$ from the two visual inputs.

\subsection{Baseline Adaptation for CASS}
DAVIS-Flow~\cite{huang2025davisflow} originally processes a single visual input and predicts only the corresponding target source.
To adapt it to the CASS setting, we use the model to predict the dialogue (DX) and effects (FX) components, and treat the residual (mixture minus predicted DX and FX) as the music (MX) estimate.

\section{Pseudo Code for Training and Inference}
\label{app:pseudo}
We provide the pseudo-code for training and sampling processes of AV-CASS in Alg.~\ref{alg:avcass_train} and Alg.~\ref{alg:avcass_infer} in this technical appendix. $\text{CAT}(\cdot)$ in the provided algorithms indicates channel-wise concatenation operation.

\section{Statistics of the Training Data}
\label{app:stats_avdnr_data}
\begin{figure}[h]
  \centering
   \includegraphics[width=\linewidth]{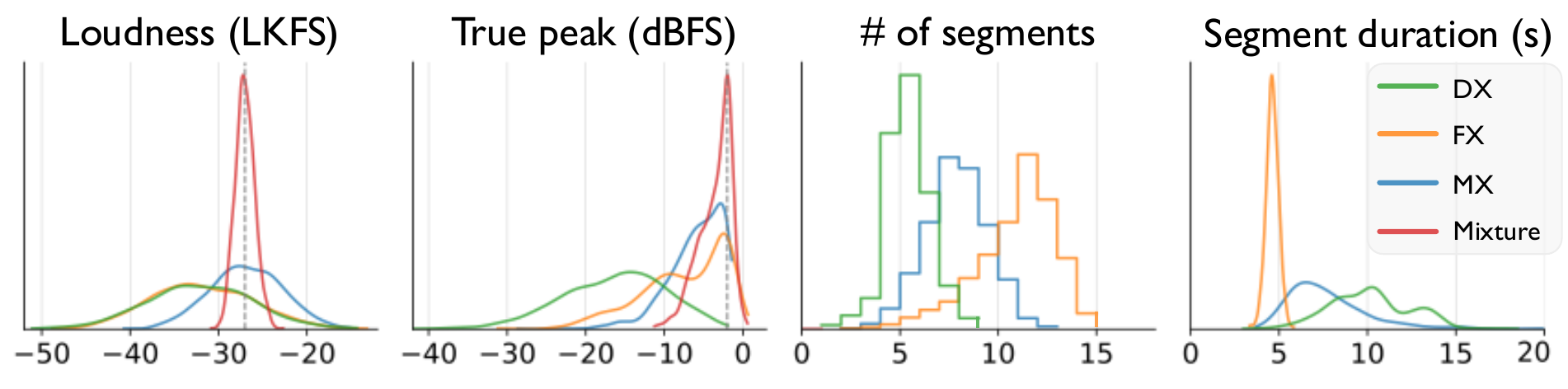}
   \caption{Distributions of the synthesized training data for loudness in LKFS, true peak in dBFS, number of segments for each stem, and the duration of each segment in seconds.
   } 
   \label{fig:avdnr_stats}
   \vspace{-4mm}
\end{figure}

We synthesize 10k training samples, each 60 seconds long, following the pipeline described in \textit{Training Data Construction Pipeline} in the main paper. Fig.~\ref{fig:avdnr_stats} presents the statistics of the training data from our training data construction pipeline. The loudness distribution of mixture audio shows an average loudness of –27 LKFS with a true peak value of –2 dBFS, ensuring that the synthesized cinematic audio closely resembles the loudness characteristics of real-world film audio~\cite{watcharasupat2024remastering}.
We also report the number of segments per stem type contained in a 60-second mixture, along with the duration distribution of individual segments for each stem.

\input{tables/algorithm_pytorch}

\section{Details of MOS evaluation}
\label{app:real-world-mos}

We conducted a Mean Opinion Score (MOS) test as a user study to evaluate the perceptual quality of audio separated by various methods. This assessment was performed on both real-world audio-visual samples from movies and synthetic samples from the AVDnR dataset in separate studies. The primary motivation for including MOS on real-world data is the absence of ground-truth (GT) references, which makes objective evaluation nearly impossible. In contrast, the AVDnR evaluation helps reveal how well objective metrics align with human judgment, as it includes clean, separated source tracks. For real-world samples, we compare with the existing CASS methods, while for AVDnR, we also include the music source separation models, \ie, HDemucs, HTDemucs, and MSDM.

\subsection{Sample Selection}
For real-world samples, we selected 30 audio-visual segments from the Condensed Movies dataset~\cite{bain2020condensed}.
Each segment was 4 to 10 seconds long and was chosen to contain all three target separation stems: DX (dialogue), FX (sound effects), and MX (music).

For AVDnR, we select 30 5-second-long audio clips from synthetic AVDnR test set.
Since ground-truth stems are available, we show the GT audio for the specified stem prior to evaluating each sample. This allows listeners to score each sample based on the comparison with the given GT. 

\subsection{Test Procedure}
\begin{figure}[t]
  \centering
   \includegraphics[width=0.95\linewidth]{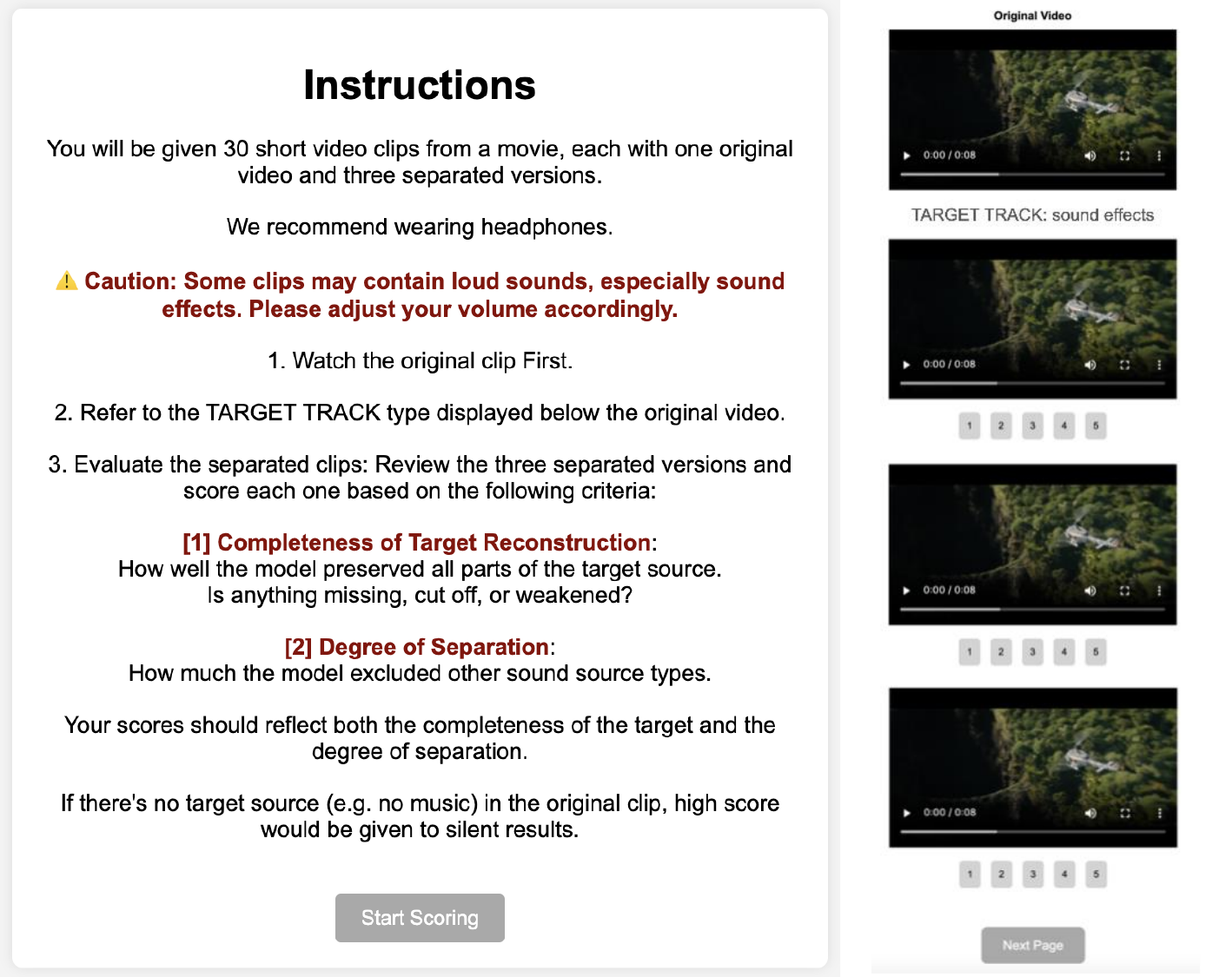}
   \caption{Instructions and scoring interface for the MOS test on real-world movie clips.
   } 
   \label{fig:mos_movies_screenshot}
\end{figure}

\begin{figure}[t]
  \centering
   \includegraphics[width=\linewidth]{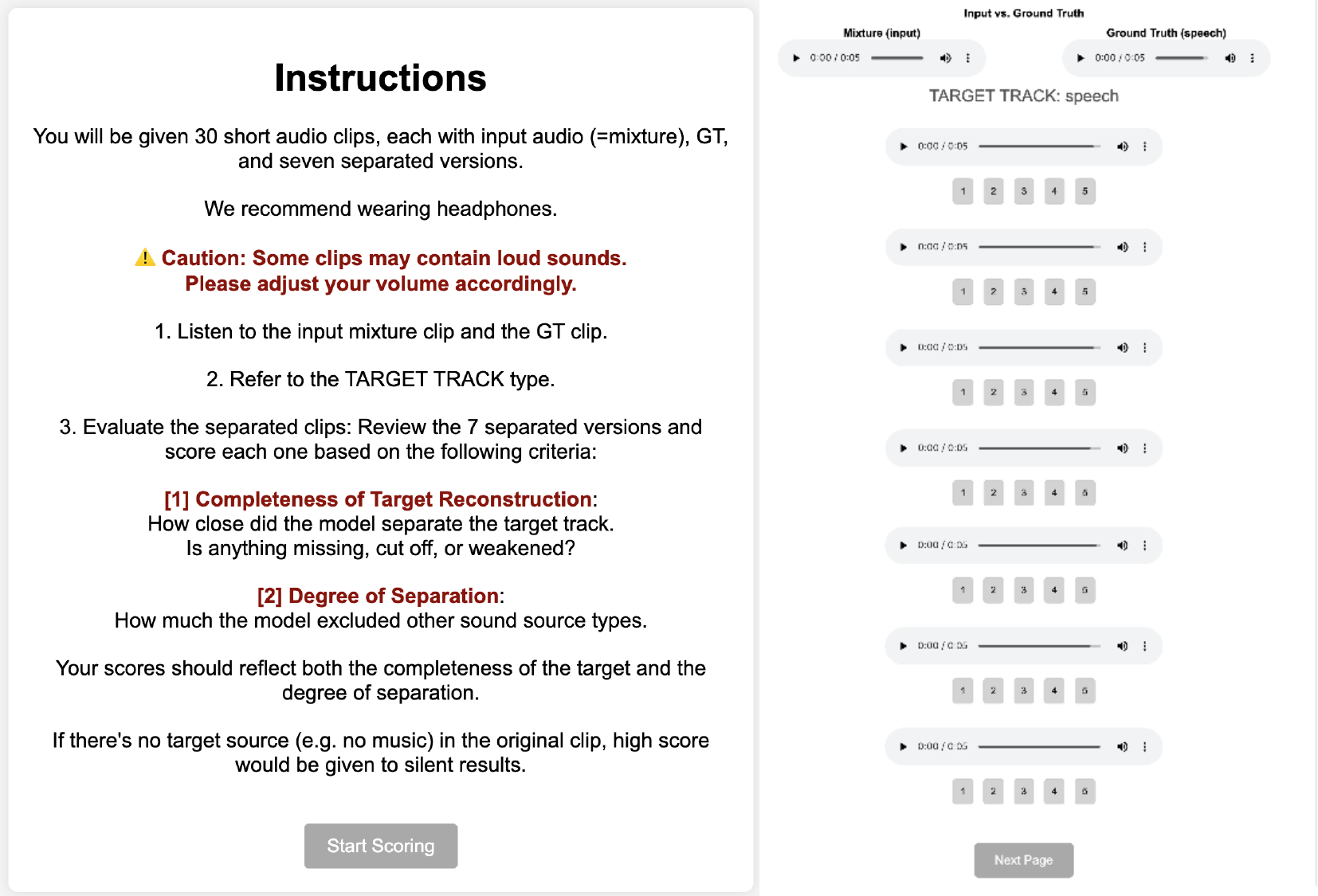}
   \caption{Instructions and scoring interface for the MOS test on AVDnR dataset.
   } 
   \label{fig:mos_avdnr_screenshot}
\end{figure}

All audio clips were volume-normalized to minimize loudness bias. Participants listened to samples in randomized order, and each clip was evaluated only once.
The instructions shown to the participants before the start of the test are displayed on the left side of Fig.~\ref{fig:mos_movies_screenshot} for real-world samples, and the left side of Fig.~\ref{fig:mos_avdnr_screenshot} for AVDnR samples. After carefully reading the instructions, the participants start the evaluation.

For real-world samples, as shown on the right side of Fig.~\ref{fig:mos_movies_screenshot}, each participant first watches the original video on the top of the scoring page, which helps participants to understand the overall audio scene and the different layers of sound present. Then, there is a text note ``TARGET TRACK'' which is the target sound type that the participant should focus on in the current evaluating sample.
Finally, the separation results from the three CASS models are presented, and each separated audio sample is synchronized with the video and placed in a randomized order. Participants then evaluate each sample based on the quality of the separated track for the given target. This setup allows participants to assess how well each model separates the target stem in the context of the original audio-visual scene.

For AVDnR samples, as shown on the left side of Fig.~\ref{fig:mos_avdnr_screenshot}, participants first listen to the mixture audio, the GT audio for the target track, and then evaluate the separation outputs from six different models. Note that we also put the GT sample into the six evaluation samples to get the score of the GT sample, which can serve as a perceptual upper bound when evaluating model performance. 
Therefore, as shown on the right side of Fig.~\ref{fig:mos_avdnr_screenshot}, participants are asked to score a total of seven samples. Although the GT is provided before each evaluation, it is not identified during the scoring of the seven samples in random order to avoid bias in the evaluation.

\subsection{Participants and Scoring}
The real-world and AVDnR samples were evaluated by 27 and 15 participants, respectively. Based on their responses, average MOS scores with 95\% confidence intervals are reported. The results for the real-world and AVDnR evaluations are shown in Tab.~\ref{tab:mos} and Tab.~\ref{tab:cass_MOS_only} of the main paper, where our method is preferred by listeners. The MOS for the ground-truth tracks of AVDnR was 4.5, reflecting the general scoring tendency of the participant group. This value serves as a perceptual upper bound for interpreting the MOS of model outputs on the AVDnR dataset. There is no perceptual upper bound for the real-world experiment, as ground-truth references are not available for real-world samples.

\section{Metrics}
\label{app:metrics}
\subsection{Standard Quality Metrics}
We evaluate separation quality using standard objective and perceptual metrics, including FAD, KL divergence, SI-SDRi, and PESQ. These metrics capture signal fidelity, perceptual quality, and distributional similarity to reference audio. We evaluate FAD and KL divergence using the official evaluation code\footnote{\url{https://github.com/haoheliu/audioldm_eval}} provided by AudioLDM~\cite{liu2023audioldm}. 

\paragraph{Fr\'echet Audio Distance (FAD).}
Fr\'echet Audio Distance (FAD)~\cite{kilgour2019fad} is a metric to measure the distance between the generated outputs and the ground truth, similar to Fr\'echet Inception Distance (FID)~\cite{heusel2017gans} in the image domain. 
We use the VGGish model pretrained on YouTube-100M to extract embeddings from both the separated and ground truth audio clips of AVDnR. FAD is then computed as the Fréchet distance between the two sets of embeddings. A lower FAD score indicates that the generated audio is more plausible and perceptually closer to real-world clean recordings. 

\paragraph{Kullback–Leibler (KL) divergence.}
Kullback-Leibler (KL) divergence measures how much the predicted distribution is different from a true distribution. 
The KL divergence computes a pairwise KL divergence between the extracted feature from the separated audio and ground truth audio, and reports the average across all the evaluation set.
Specifically, we use the PANNs~\cite{kong2020panns} model pretrained on AudioSet for large-scale audio classification to extract class probability distributions from both the generated and reference audio. 
A lower KL divergence suggests that the model output shares similar semantic content with the ground truth, implying better preservation of the original sound concepts.

\input{tables/CASS_trainOurs_testAVDnR_detailed}

\paragraph{Scale-Invariant Signal-to-Distortion Ratio improvement (SI-SDRi).}
SI-SDRi measures the quality improvement of a separated signal compared to the input signal while considering the scale-invariance of audio signals. A higher SI-SDRi indicates better separation performance, with 0 dB signifying no improvement over the input.

The scale-invariant signal-to-distortion ratio (SI-SDR)~\cite{le2019sdr} is defined as:
\begin{equation}
    \mathrm{SI}\text{-}\mathrm{SDR}(\bm{x},\hat{\bm{x}})=10 \mathrm{log} _{10} \frac{\left \| \alpha \bm{x}    \right \|^{2} +\epsilon  }{\left \|  \alpha \bm{x}-\hat{\bm{x}}\right \|^{2}+\epsilon  },
    \label{eq:SI_SDRi}
\end{equation}
where $\bm{x}$ and $\hat{\bm{x}}$ denote the ground-truth and estimated sources, respectively. The optimal scaling factor $\alpha$ for the target signal is defined as:
\begin{equation} 
\alpha = \frac{\bm{x}^{\top} \hat{\bm{x}} + \epsilon}{\|\bm{x}\|^{2} + \epsilon}, 
\end{equation}
with $\epsilon = 9.76562 \times 10^{-4}$ to ensure numerical stability.
Given the ground-truth audio signal $\bm{x}$, input mixture audio $\bm{y}$, and the estimated source $\hat{\bm{x}}$, SI-SDRi is calculated as the improvement over the mixture baseline:

\begin{equation} 
\mathrm{SI}\text{-}\mathrm{SDRi} = \mathrm{SI}\text{-}\mathrm{SDR}(\bm{x}, \hat{\bm{x}}) - \mathrm{SI}\text{-}\mathrm{SDR}(\bm{x}, \bm{y}). 
\end{equation}

\paragraph{Perceptual Evaluation of Speech Quality (PESQ).}
We use PESQ~\cite{pesq} to evaluate the perceptual quality of the separated speech track, which is widely used in speech enhancement tasks. PESQ uses psychoacoustic modeling to measure perceived distortions between the ground truth and predicted signals. The score ranges from –0.5 to 4.5, where higher is better. 

\subsection{Wrong Placement Ratio (WPR)}
In this paper, we introduce a new metric, Wrong Placement Ratio (WPR), to assess the presence of incorrectly placed sounds in each separated stem. This metric does not require ground-truth reference separated tracks. It uses a pretrained frame-wise sound event detection (SED) model, PANNs~\cite{kong2020panns}, to detect segments of residual or misplaced components of other stems in each target stem.
The SED operates with a frame resolution of 10 ms and produces per-frame activation probabilities for 527 sound event categories. These fine-grained classes are manually grouped into three broad categories, \ie speech, sound effects, and music, based on their semantic labels and auditory characteristics. {The grouping taxonomy can be found on the attached "class\_labels\_with\_main\_class.csv".}

For each separated track, we apply the SED model and convert the resulting probability matrix into a binary matrix $P \in \{0,1\}^{T \times C}$, where $T$ is the number of frames and $C = 3$ is the number of merged classes. An element $P_{t,c} = 1$ indicates that class $c$ is predicted to be active at frame $t$. 
A fixed threshold of 0.25 is applied to the SED probabilities to obtain the binary decisions.

We define WPR for each track to quantify the proportion of frames contaminated by non-target sound classes, \ie, for the separated speech track, we measure how many sound effects and music segments were wrongly placed there. {To ensure the metric captures substantive misplacements and excludes transient artifacts, we only count non-target activations that persist for a minimum length threshold $\tau$. This threshold is empirically set to 50 frames (0.5 seconds).}
Let $\text{target}_i$ denote the expected class label for separated track $i$, where $i \in {1, 2, 3}$ corresponds to speech, sound effects, and music, respectively.
{We define a thresholded binary prediction matrix $\hat{P}$ where $\hat{P}_{t, c} = 1$ only if the prediction for class $c$ at time $t$ belongs to a contiguous block of at least $\tau$ activated frames.}
We calculate the WPR based on non-silent frames, which are frames with at least one class activated in the original prediction matrix $P$. The WPR for separated track $i$ is computed as follows:
\begin{equation}
\text{WPR}_i = \frac{1}{T_i} \sum_{t=1}^{T_i} 1 \left[\sum_{c \neq \text{target}_i} \hat{P}_{t, c} > 0\right]
\end{equation}
\noindent where $T_i$ is the sum of non-silent frames in track $i$, and the indicator function returns $1$ if any non-target class is active in frame $t$. We exclude silent frames from this calculation because they trivially yield a WPR of zero, which does not reflect correct separation behavior when the target class is missing entirely. This ensures that the metric captures actual contamination in tracks with meaningful content.

\section{Per-Track Objective Metrics on AVDnR}
\label{app:detailed_metrics}
Tab.~\ref{tab:cass_trainOurs_testAVDnR_detailed} presents detailed metric scores for each stem on the AVDnR dataset, along with the average values across all stems. Our model consistently outperforms other methods across all stems in terms of FAD and achieves the highest PESQ score for the DX stem. It also achieves the best KL divergence scores for the DX and MX stems and performs comparably to the top method on the FX stem. While our model shows lower performance on the SI-SDRi metric, this is expected given the nature of generative models, which often prioritize perceptual quality over waveform-level fidelity, as discussed in the main paper. 

For the per-track WPR shown in Tab.~\ref{tab:cass_trainOurs_testAVDnR_detailed}, AV-CASS achieves the lowest average leakage (1.84\%) and sets new best results on DX (0.03\%) and MX (1.21\%). The only stem where AV-CASS is not the top performer is FX, where DAVIS-Flow~\cite{huang2025davisflow} achieves the lowest WPR (3.66\%), reflecting its strong reliance on visual object cues for transient effects. However, DAVIS-Flow performs significantly worse on DX and MX, with large leakages of 0.22\% (DX) and 32.55\% (MX), highlighting its difficulty in handling dialogue and music where audio-visual associations are weaker or more complex. In contrast, AV-CASS delivers consistently low leakage across all stems, demonstrating robust and well-balanced separation quality.

\section{Per-Track WPR Results on DnRv2 and DnRv3}
\input{tables/AO_CASS_WPR}

We presented per-track WPR results on real-world samples and the AVDnR dataset in the main paper. As part of the main paper, we also showed quantitative results on the DnRv2 and DnRv3 datasets using audio-only CASS methods in Tab.~\ref{tab:cass_trainOurs_testALL} of the main paper, but only with average WPR scores. Here in the appendix, we provide the per-track WPR results for DnRv2 and DnRv3 in Tab.~\ref{tab:wpr_trainOurs_testALL}. Our model consistently outperforms others, showing particularly large improvements on the FX stem. While it slightly underperforms on the DX track of DnRv3, the performance gap is minimal (0.05\%). These results demonstrate our model’s flexibility and generalizability in audio-only setup, achieving accurate separation with minimal cross-track confusion.

\section{Qualitative Results on AVDnR}
\label{app:supple_additional_results}

\begin{figure}[ht]
  \centering
   \includegraphics[width=\linewidth]{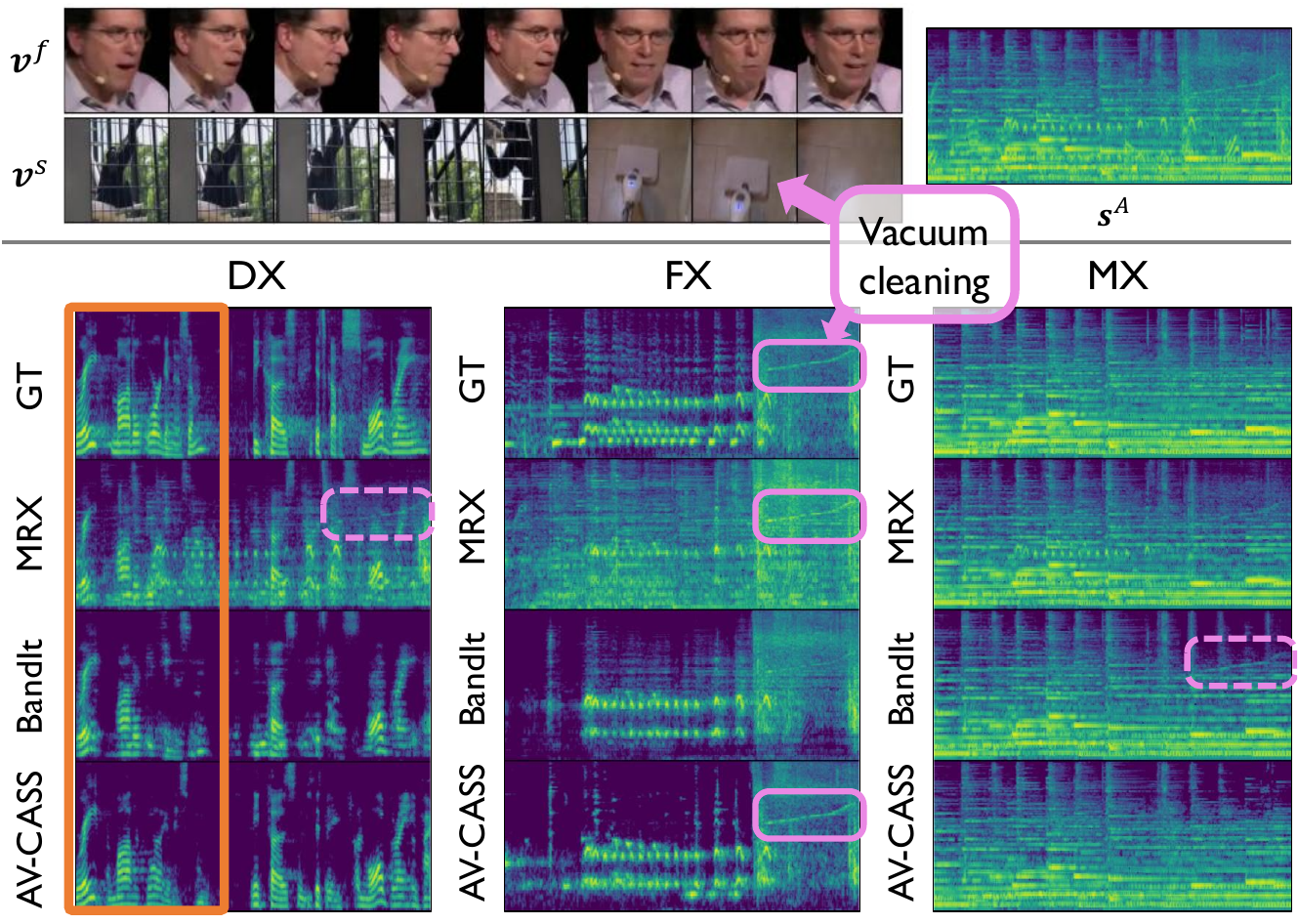}
   \caption{\textit{Comparison of spectrograms from MRX, BandIt, and AV-CASS (ours) on AVDnR.} Facial frames $\bm v^f$, scene frames $\bm v^s$, and the input audio spectrogram $\bm s^A$ shown at the top. Ground truth audio spectrogram is on top of the models' results for each stem. The orange box in the DX stem highlights that our model reconstructs speech most accurately compared to others. Pink solid boxes indicate the presence of vacuum cleaning in $\bm{v}^s$ and the spectrograms of the FX stem, while dotted boxes indicate incorrectly placed vacuum cleaning sounds.}
   \label{fig:res_spec_vacuum}
\end{figure}

For qualitative assessment, we present spectrograms from our model and existing CASS methods in Fig.~\ref{fig:res_spec_vacuum}. 
Unlike previous audio-only approaches, our model consistently isolates sounds with minimal misclassification. 
Notably, in the orange-boxed area, AV-CASS accurately reconstructs speech even under challenging conditions with significant overlap from music and sound effects, leveraging facial cues for improved separation.
Additionally, the vacuum cleaning sound, indicated by the pink arrow on the FX track and associated with a visible vacuum cleaner in $\bm{v}^{s}$, is correctly separated as FX by our model. In contrast, other methods incorrectly place it on music or speech tracks. 

\begin{figure*}[ht]
  \centering
   \includegraphics[width=0.9\linewidth]{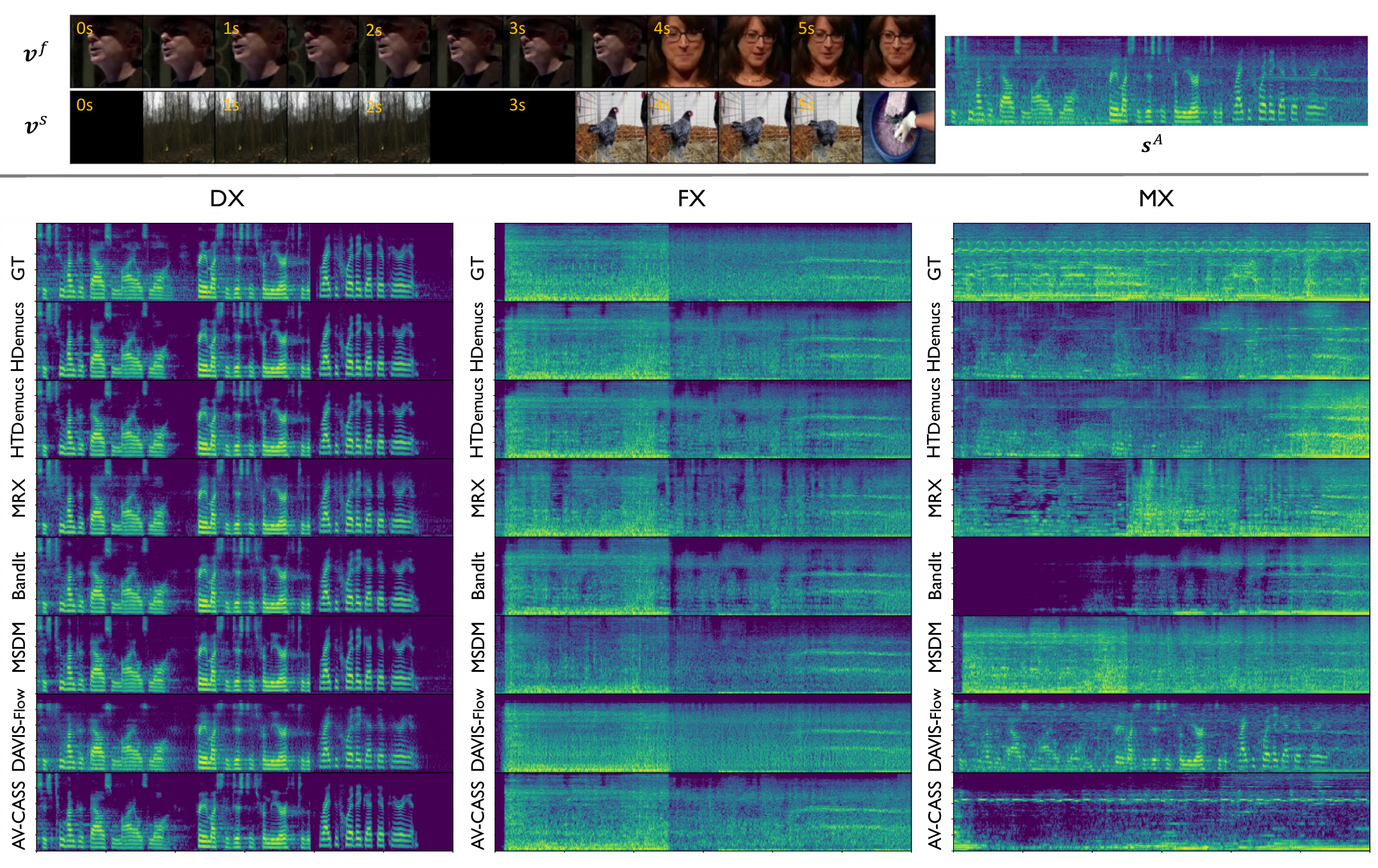}
   \caption{Comparison of the separation results of our model and other methods on AVDnR dataset.}
   \label{fig:AVDnR_sample_parts_0_7}
\end{figure*}

\begin{figure*}[ht]
  \centering
   \includegraphics[width=0.9\linewidth]{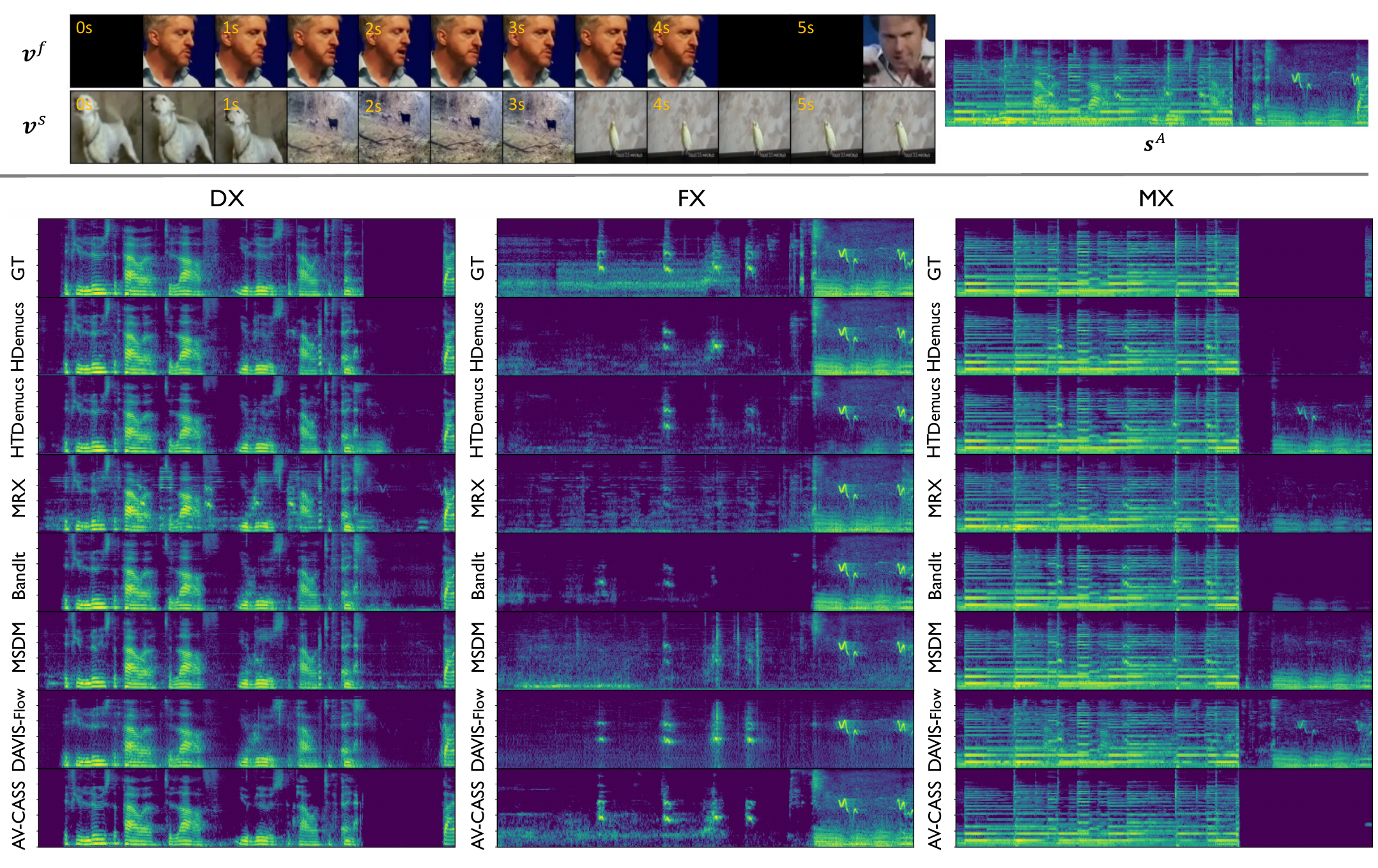}
   \caption{Comparison of the separation results of our model and other methods on AVDnR dataset.}
   \label{fig:AVDnR_sample_parts_0_12}
\end{figure*}

\begin{figure*}[ht]
  \centering
   \includegraphics[width=0.9\linewidth]{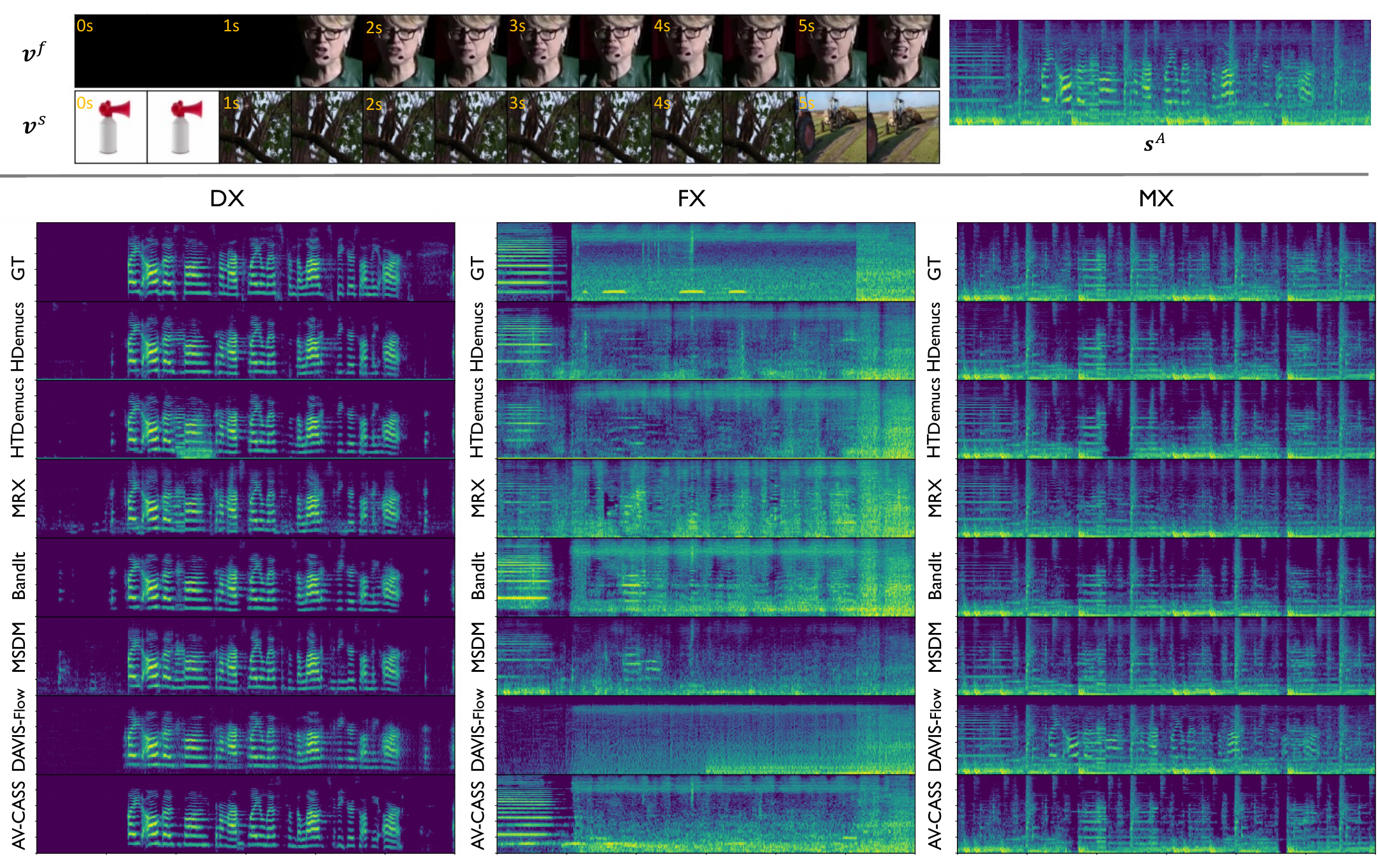}
   \caption{Comparison of the separation results of our model and other methods on AVDnR dataset.}
   \label{fig:AVDnR_sample_parts_0_22}
\end{figure*}

\begin{figure*}[ht]
  \centering
   \includegraphics[width=0.9\linewidth]{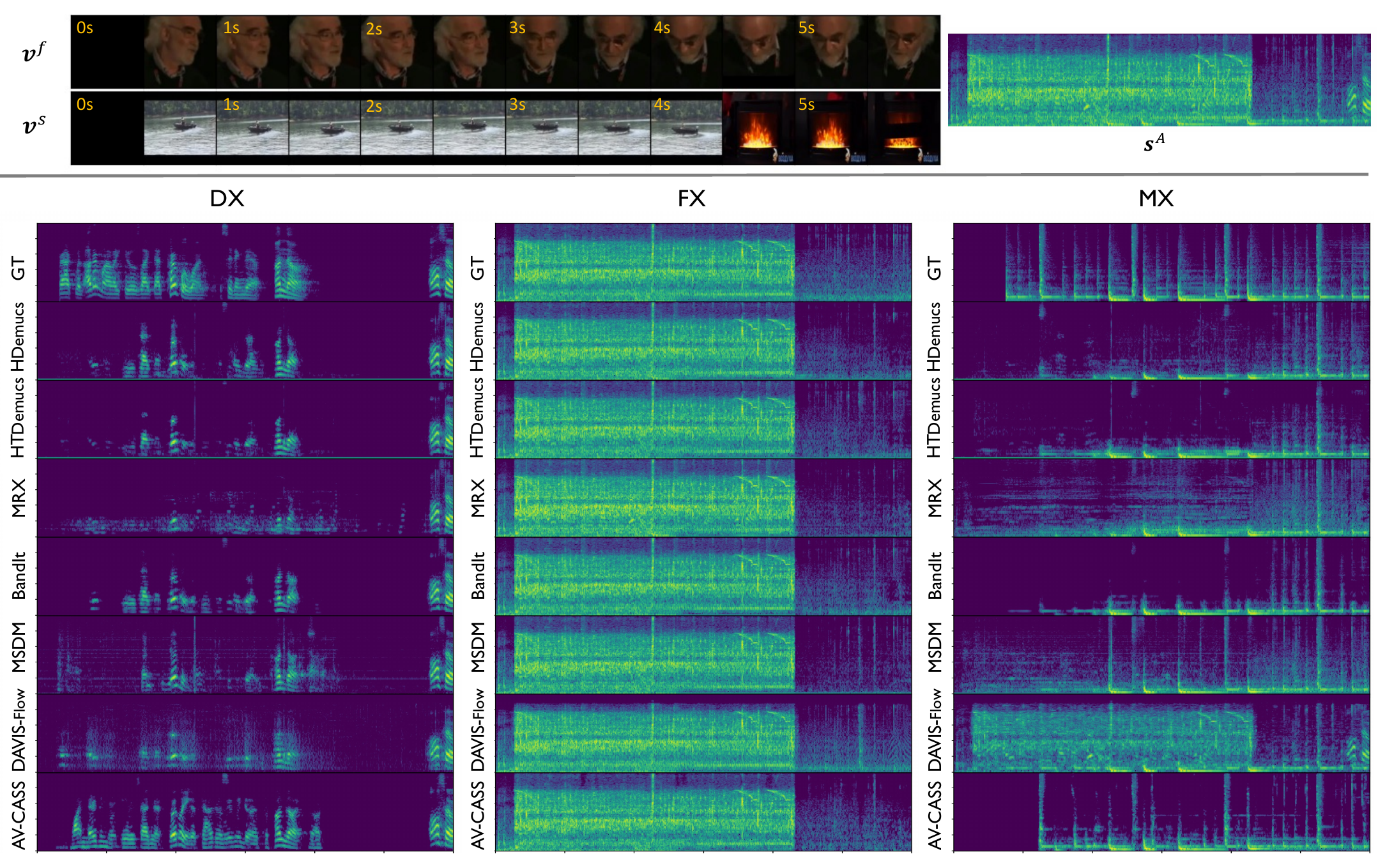}
   \caption{Comparison of the separation results of our model and other methods on AVDnR dataset.}
   \label{fig:AVDnR_sample_parts_0_53}
\end{figure*}

\begin{figure*}[ht]
  \centering
   \includegraphics[width=0.9\linewidth]{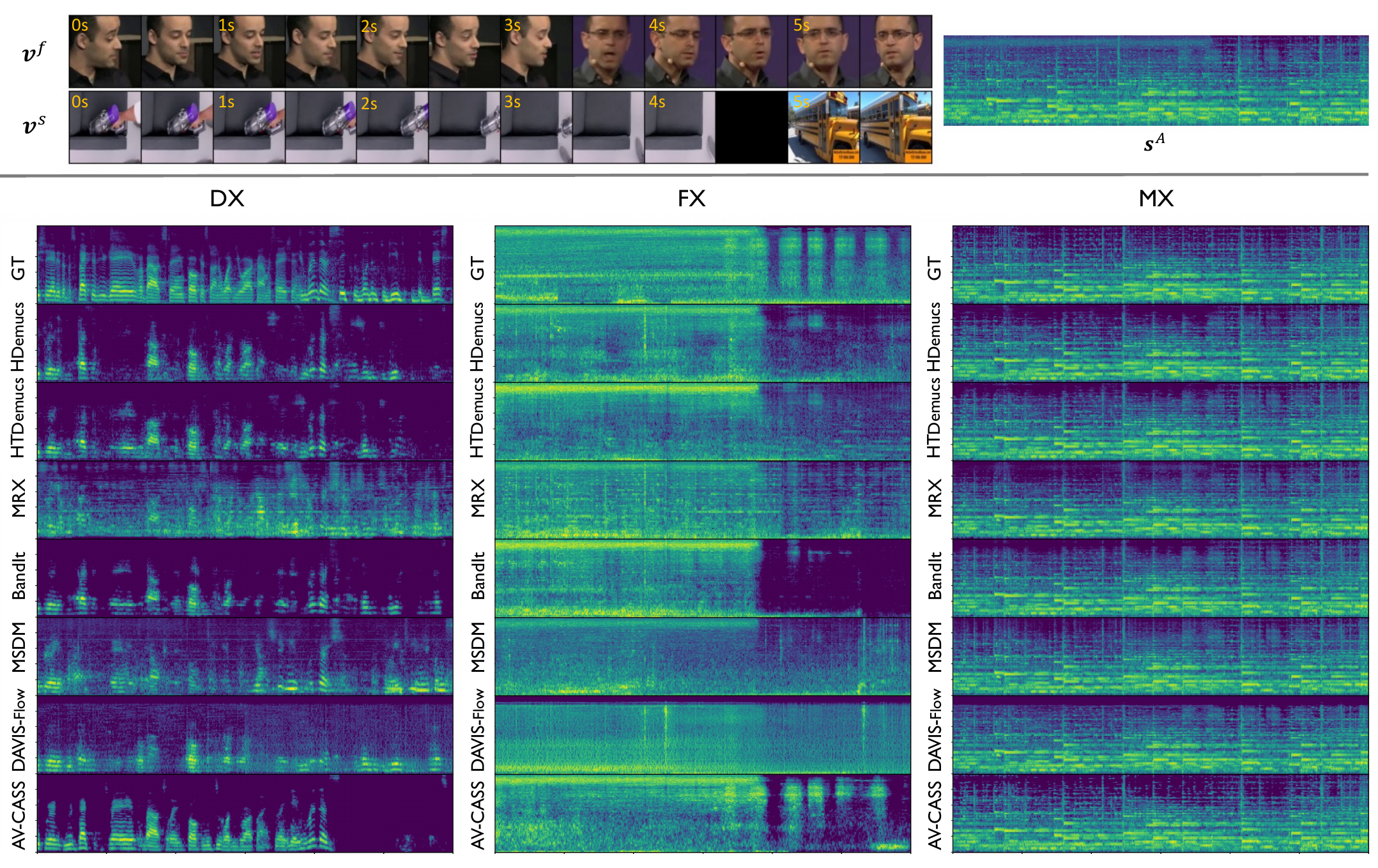}
   \caption{Comparison of the separation results of our model and other methods on AVDnR dataset.}
   \label{fig:AVDnR_sample_parts_0_75}
\end{figure*}

Furthermore, we present additional comparisons on the AVDnR dataset in Fig.~\ref{fig:AVDnR_sample_parts_0_7} through Fig.~\ref{fig:AVDnR_sample_parts_0_75}.
These figures compare our audio-visual model with existing music source separation and CASS models, including Hybrid Demucs (HDemucs)~\cite{defossez2021hybrid}, HT Demucs~\cite{rouard2022hybrid}, MSDM~\cite{mariani2024multisource}, MRX~\cite{petermann2022cocktail}, and BandIt~\cite{watcharasupat2023bandit}. 
The visualizations highlight the effectiveness of our approach in accurately separating audio components into each corresponding track. 

\begin{figure*}[!ht]
  \centering
   \includegraphics[width=0.9\linewidth]{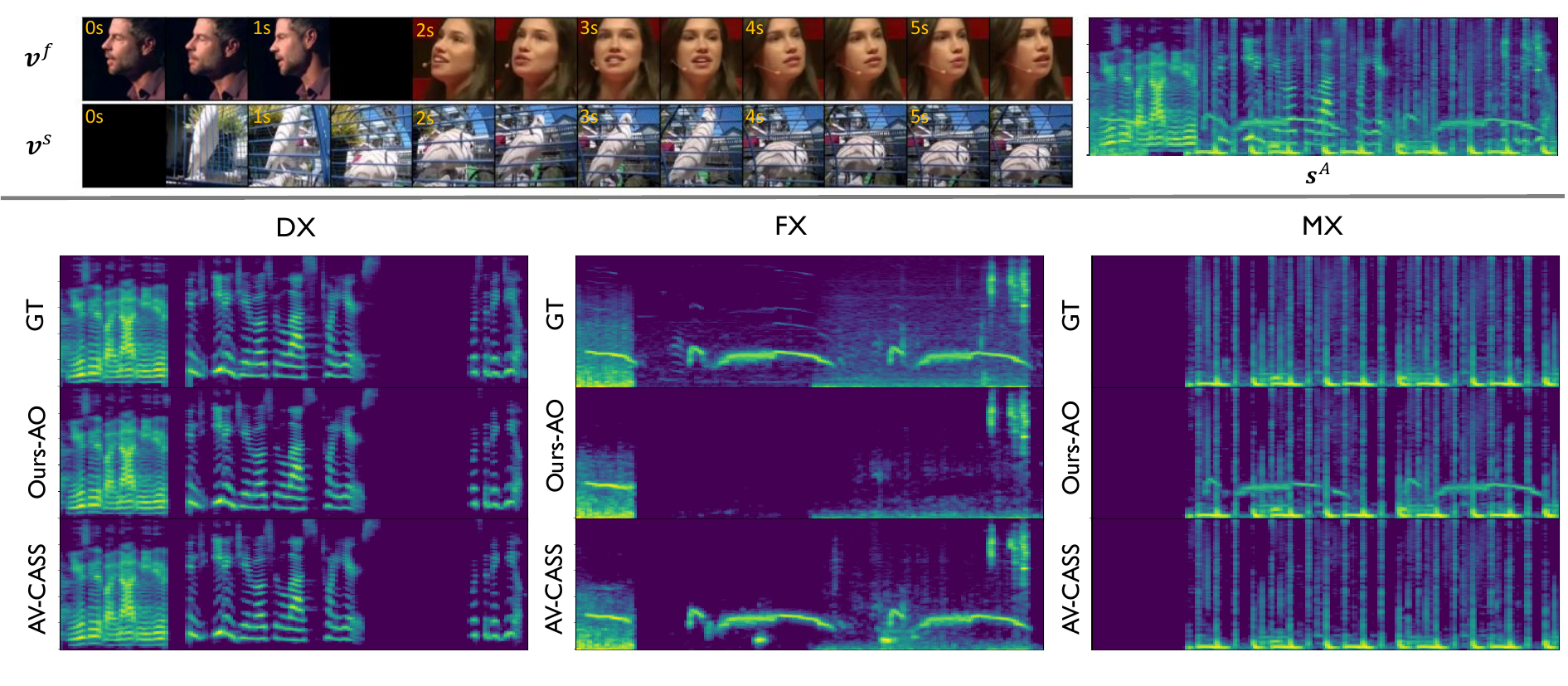}
   \caption{Spectrogram comparison of separated outputs from our audio-only model (Ours-AO), audio-visual model (AV-CASS), and ground truth (GT) on AVDnR.}
   \label{fig:res_spec_bird_cooing}
\end{figure*}

\begin{figure*}[ht]
  \centering
   \includegraphics[width=0.9\linewidth]{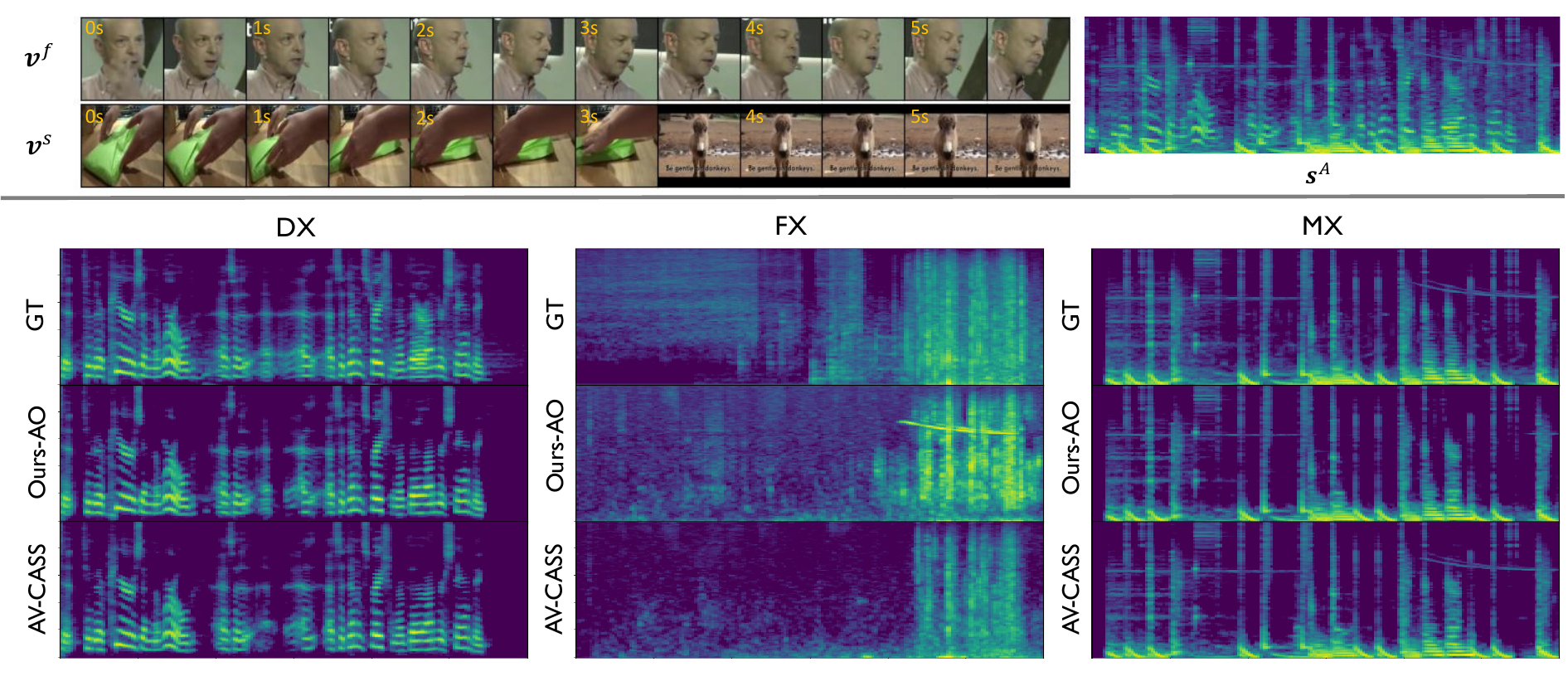}
   \caption{Spectrogram comparison of separated outputs from our audio-only model (Ours-AO), audio-visual model (AV-CASS), and ground truth (GT) on AVDnR.}
   \label{fig:AVDnR_sample_parts_6_67}
\end{figure*}

Similar to Sec.~\ref{sec:eval_on_avdnr} in the main paper, we include additional results comparing our audio-only and audio-visual models in Fig.~\ref{fig:res_spec_bird_cooing} to Fig.~\ref{fig:AVDnR_sample_parts_6_67}, demonstrating the effectiveness of visual cues in improving separation quality.

\section{Failure Case Analysis}
Among the real-world sample results, we observed several failure cases for our model.

\noindent\textbf{Difficulty with non-verbal vocalizations.}
First, our model showed difficulty in handling non-verbal vocalizations, such as screaming and laughter. Since both our training data and previous DnR datasets included a limited amount of data for screaming or laughter within the DX stem, the model struggles to consistently separate these sounds into the designated speech track. This issue arises from the ambiguity of the class boundary and the lack of such data in the training set. Given the data-centric nature of the problem, other existing CASS models also share this limitation. This specific issue has recently been addressed in audio-only CASS by introducing a new, dedicated dataset \cite{hasumi2025dnr}. We plan to resolve this on the audio-visual dataset side, or by pretraining the model on these audio-only datasets to provide the necessary guidance.

\noindent\textbf{Generative artifacts from loud sound effects.}
Second, when encountering a loud sound effect that occupies a wide range of frequency bands (\eg a loud engine sound), our model sometimes generates speech-like artifacts even in the absence of actual speech. This is likely caused by the inherent characteristics of generative models which may over-generalize complex noise patterns as vocal cues. In future work, we will investigate methods to explicitly penalize such generative artifacts during training and inference.

\noindent\textbf{Confusion between music and synchronized SFXs.}
The final failure case observed occurs specifically in contexts like musical films or movie trailers. When the music's rhythm and the SFX timing are edited in perfect synchronization, the model shows a tendency to confuse the music and SFX. It frequently leads to imperfect separation, where the output streams are not cleanly isolated and contain components of the other sound source.

%% file: tables/algorithm_pytorch.tex
\renewcommand{\lstlistingname}{Algorithm}

\begin{figure*}[t]
\centering
\begin{minipage}{\textwidth}
\begin{lstlisting}[style=PyTorchStyle, caption={PyTorch-style pseudocode for AV-CASS training.}, label={alg:avcass_train}]
# --- Setup ---
# VFE_model (u_theta): torch.nn.Module
# optimizer: torch.optim.Optimizer
# criterion: torch.nn.MSELoss (L2 loss)
# D: torch.utils.data.DataLoader
# visual_fusion: Function/Module for C_V = F_theta(E^f(v_f) || E^s(v_s))

VFE_model.train()

for s_A, v_f, v_s, s_DX, s_FX, s_MX in D:
    # 1. Compute fused visual condition vector c_V
    c_V = visual_fusion(v_f, v_s) 

    # 2. Sample time step t (according to t = 1/(1 + exp(-s)), s ~ N(0, 1))
    s = torch.randn_like(s_DX)
    t = torch.sigmoid(s) 
    
    # 3. Sample initial noise x_0 and concatenate target sources x_1
    x_0 = torch.randn_like(s_DX) 
    x_1 = torch.cat([s_DX, s_FX, s_MX], dim=1) # Target clean sources
    
    # 4. Compute the point at time step t: x_t = (1-t)x_0 + t*x_1
    # Adjust t's shape for proper broadcasting across audio dimensions
    x_t = (1 - t) * x_0 + t * x_1
    
    # 5. Model Input and Prediction
    input_cat = torch.cat([x_t, s_A], dim=1)
    u_pred = VFE_model(input_cat, t, c_V)
    
    # 6. Compute Target Vector Field: u_target = x_1 - x_0
    u_target = x_1 - x_0
    
    # 7. MSE flow-matching loss: L = ||u_pred - u_target||^2_2
    loss = criterion(u_pred, u_target)

    # 8. Update Weights
    optimizer.zero_grad()
    loss.backward()
    optimizer.step()
\end{lstlisting}
\end{minipage}
\end{figure*}

\begin{figure*}[t]
\centering
\begin{minipage}{0.95\textwidth}
\begin{lstlisting}[style=PyTorchStyle, caption={PyTorch-style pseudocode for AV-CASS inference.}, label={alg:avcass_infer}]
# --- Requirements ---
# VFE_model: Trained torch.nn.Module (u_theta)
# s_A: Mixture spectrogram
# v_f, v_s: Visual frames
# N: Sampling time steps (e.g., N=100)

@torch.no_grad()
def inference_av_cass(VFE_model, s_A, v_f, v_s, N):
    VFE_model.eval()
    
    # 1. Compute step size and condition vector
    eta = 1.0 / N
    c_V = visual_fusion(v_f, v_s) 
    
    # 2. Initialize x_t (x_0) with Gaussian noise
    batch_size, channels_A, T, F = s_A.shape
    required_channels = 3 * channels_A # Assuming 3 sources: DX, FX, MX
    x_t = torch.randn(batch_size, required_channels, T, F, device=s_A.device)
    
    # 3. Euler Integration Loop (n = 1 to N)
    for n in range(1, N + 1):
        t = n / N # Normalized time step
        
        # Compute VFE direction: u = u_theta(CAT(x_t, s_A), t, c_V)
        input_cat = torch.cat([x_t, s_A], dim=1)
        u = VFE_model(input_cat, t_tensor, c_V)
        
        # Euler Step: x_{t+eta} = x_t + eta * u
        x_t = x_t + (eta * u)
        
    # 4. Return Separated Sources
    # The final x_t is the estimate of x_1 (CAT(s_DX, s_FX, s_MX))
    pred_s_DX, pred_s_FX, pred_s_MX = torch.chunk(x_t, chunks=3, dim=1)
    
    return pred_s_DX, pred_s_FX, pred_s_MX
\end{lstlisting}

\end{minipage}
\end{figure*}

%% file: tables/CASS_trainOurs_testAVDnR_detailed.tex
\begin{table*}[ht]
\centering
\resizebox{1.0\linewidth}{!}{
\begin{tabular}{lcc|ccccccccccccccccc}
\toprule
\toprule
  \multirow{2}{*}{\textbf{Method}} 
  & \multirow{2}{*}{\textbf{A-V}} 
  & \multirow{2}{*}{\textbf{P/G}} 
  & \multicolumn{4}{c}{\textbf{FAD ($\downarrow$)}} 
  & \multicolumn{4}{c}{\textbf{KL ($\downarrow$)}} 
  & \multicolumn{4}{c}{\textbf{SI-SDRi [dB] ($\uparrow$)}} 
  & \multicolumn{4}{c}{\textbf{WPR [\%] ($\downarrow$)}} 
  & {\textbf{PESQ ($\uparrow$)}} \\
  \cmidrule(lr){4-7} \cmidrule(lr){8-11} \cmidrule(lr){12-15} \cmidrule(lr){16-19} \cmidrule(lr){20-20}
      & & 
      & \textbf{Avg.}& \textbf{DX} & \textbf{FX} & \textbf{MX}  
      & \textbf{Avg.} & \textbf{DX} & \textbf{FX} & \textbf{MX}  
      & \textbf{Avg.} & \textbf{DX} & \textbf{FX} & \textbf{MX}  
      & \textbf{Avg.} & \textbf{DX} & \textbf{FX} & \textbf{MX} 
      & \textbf{DX} \\
\midrule

Hybrid Demucs~\cite{defossez2021hybrid} & \ding{55} & P
& 2.05 & \underline{1.66} & \underline{1.34} & 3.16 
& \underline{1.03} & 0.99 & \underline{0.95} & \underline{1.14}
& \underline{13.57} & \underline{12.75}& \underline{13.59} & \underline{14.38} 
& 5.24 & 0.12 & 12.67 & 2.93
& \underline{2.16}  \\

HT Demucs~\cite{rouard2022hybrid} & \ding{55} & P
& 2.08 &  1.97 & 1.62 & 2.65 
& 1.06 & 1.09& \textbf{0.92} & 1.17
& 13.41& 12.68 & 13.33 & 14.21 
& 9.23 & 0.15 & 19.97 & 7.57
& 2.06  \\

MRX~\cite{petermann2022cocktail} & \ding{55} & P
& 3.47 & 2.67 & 2.02 & 4.01 
& 1.67 & 2.00& 1.02& 1.97
& 10.60 &11.23& 11.34& 12.31
& 14.91 & 2.18 & 37.60 & 4.94
& 1.89  \\

BandIt~\cite{watcharasupat2023bandit} & \ding{55} & P
& 2.15 & 2.38 & 2.96 & \underline{1.11}
& 1.14 & 0.82& 1.19& 1.41
& \textbf{14.40} &\textbf{13.33}& \textbf{14.39}& \textbf{15.48}
& 4.65 & 0.04 & 12.46 & 1.46
& 2.15  \\

MSDM~\cite{mariani2024multisource} & \ding{55} & P
& 2.90 & 3.84 & 3.03 & 3.56 
& 1.70 & 2.47& 0.97& 1.66
& 11.63 &11.48 & 9.38 & 10.94 
& 5.65 & 1.65 & 11.49 & 3.82
& 2.12  \\

DAVIS-Flow~\cite{huang2025davisflow} & \ding{51} & G
& 5.94 & 6.72 & 3.75 & 7.35 
& 1.64 & 1.50 & 1.48 & 1.94 
& 9.25 & 9.44 & 10.96 & 7.35 
& 12.14 & 0.22 & \textbf{3.66} & 32.55
& 1.94     \\

\textbf{AV-CASS (Ours)} & \ding{51} & G
& \textbf{0.85} & \textbf{0.67} & \textbf{0.89} & \textbf{0.98} 
& \textbf{0.93} & \textbf{0.64}  & 1.05 & \textbf{1.09} 
& 12.32 & 10.93 & 12.33 & 13.71 
& \textbf{1.84} & \textbf{0.03} & \underline{4.29} & \textbf{1.21}
& \textbf{2.26}     \\

\midrule
\textcolor{gray}{Ours (Audio-only)} & \ding{55} & G
& \underline{\textcolor{gray}{1.63}} & \textcolor{gray}{2.13} & \textcolor{gray}{1.61} & \textcolor{gray}{1.15} 
& \textcolor{gray}{1.15} & \underline{\textcolor{gray}{0.71}} & \textcolor{gray}{1.37}& \textcolor{gray}{1.38}
& \textcolor{gray}{12.23} & \textcolor{gray}{10.94} & \textcolor{gray}{12.24} & \textcolor{gray}{13.52} 
& \underline{\textcolor{gray}{2.01}} & \underline{\textcolor{gray}{0.03}} & \textcolor{gray}{4.65} & \underline{\textcolor{gray}{1.34}}
& \textcolor{gray}{2.08} \\

\bottomrule
\end{tabular}
}
\caption{\textit{Detailed objective metric results on AVDnR for each stem.} All models are trained on our training data. A-V refers to audio-visual, P refers to predictive model, and G refers to generative model. }
\label{tab:cass_trainOurs_testAVDnR_detailed}
\vspace{-4mm}
\end{table*}

%% file: tables/AO_CASS_WPR.tex
\begin{table}[ht]
\centering
\resizebox{\linewidth}{!}{
\setlength{\tabcolsep}{10pt}
\begin{tabular}{lcccccc}
\toprule
      & \multicolumn{3}{c}{\textbf{DnRv2}~\cite{petermann2022cocktail}} & \multicolumn{3}{c}{\textbf{DnRv3}~\cite{watcharasupat2024remastering}} \\ 
\cmidrule(lr){2-4} \cmidrule(lr){5-7}
\textbf{Method}   & \textbf{DX}  & \textbf{FX}  & \textbf{MX}  & \textbf{DX}  & \textbf{FX}  & \textbf{MX} \\
\midrule
Hybrid Demucs~\cite{defossez2021hybrid}   & 0.26 & 10.94 & 3.13 & 0.19 & 7.75 & 3.39 \\
HT Demucs~\cite{rouard2022hybrid}  & 0.26 & 21.45 & 5.71 & 0.20 & 19.67 & 5.61 \\
MSDM~\cite{mariani2024multisource}          & 2.42 & 10.20 & 3.38 & 0.18 & 8.86  & 4.41 \\
MRX~\cite{petermann2022cocktail}           & 2.00 & 41.50 & 6.19 & 3.25 & 39.73 & 6.94 \\
BandIt~\cite{watcharasupat2023bandit}        & \underline{0.24} & \underline{7.36} &  \underline{3.40} & \textbf{0.09} & \underline{6.56} & \underline{3.95} \\
\rowcolor{lightgray!25}
\textbf{Ours (Audio-Only)} & \textbf{0.12} & \textbf{3.78} & \textbf{2.48} & \underline{0.14} & \textbf{2.46} & \textbf{3.13} \\
\bottomrule
\end{tabular}
}
\caption{\textit{Wrong Placement Ratio (WPR [\%]).} Ratio of residual sounds from other tracks within each track type for the DnRv2 and DnRv3 test sets. Lower is better.}
\label{tab:wpr_trainOurs_testALL}
\end{table}